\documentclass[12pt]{article}
\usepackage{graphicx}
\usepackage{lscape}
\usepackage{citesort}

\begin{document}

\def\ds{\displaystyle}
\def\beq{\begin{equation}}
\def\eeq{\end{equation}}
\def\bea{\begin{eqnarray}}
\def\eea{\end{eqnarray}}
\def\beeq{\begin{eqnarray}}
\def\eeeq{\end{eqnarray}}
\def\ve{\vert}
\def\vel{\left|}
\def\ver{\right|}
\def\nnb{\nonumber}
\def\ga{\left(}
\def\dr{\right)}
\def\aga{\left\{}
\def\adr{\right\}}
\def\lla{\left<}
\def\rra{\right>}
\def\rar{\rightarrow}
\def\nnb{\nonumber}
\def\la{\langle}
\def\ra{\rangle}
\def\ba{\begin{array}}
\def\ea{\end{array}}
\def\tr{\mbox{Tr}}
\def\ssp{{\Sigma^{*+}}}
\def\sso{{\Sigma^{*0}}}
\def\ssm{{\Sigma^{*-}}}
\def\xis0{{\Xi^{*0}}}
\def\xism{{\Xi^{*-}}}
\def\qs{\la \bar s s \ra}
\def\qu{\la \bar u u \ra}
\def\qd{\la \bar d d \ra}
\def\qq{\la \bar q q \ra}
\def\gGgG{\la g^2 G^2 \ra}
\def\q{\gamma_5 \not\!q}
\def\x{\gamma_5 \not\!x}
\def\g5{\gamma_5}
\def\sb{S_Q^{cf}}
\def\sd{S_d^{be}}
\def\su{S_u^{ad}}
\def\ss{S_s^{??}}
\def\sbp{{S}_Q^{'cf}}
\def\sdp{{S}_d^{'be}}
\def\sup{{S}_u^{'ad}}
\def\ssp{{S}_s^{'??}}
\def\sig{\sigma_{\mu \nu} \gamma_5 p^\mu q^\nu}
\def\fo{f_0(\frac{s_0}{M^2})}
\def\ffi{f_1(\frac{s_0}{M^2})}
\def\fii{f_2(\frac{s_0}{M^2})}
\def\O{{\cal O}}
\def\sl{{\Sigma^0 \Lambda}}
\def\es{\!\!\! &=& \!\!\!}
\def\ap{\!\!\! &\approx& \!\!\!}
\def\ar{&+& \!\!\!}
\def\ek{&-& \!\!\!}
\def\kek{\!\!\!&-& \!\!\!}
\def\cp{&\times& \!\!\!}
\def\se{\!\!\! &\simeq& \!\!\!}
\def\eqv{&\equiv& \!\!\!}
\def\kpm{&\pm& \!\!\!}
\def\kmp{&\mp& \!\!\!}


\def\simlt{\stackrel{<}{{}_\sim}}
\def\simgt{\stackrel{>}{{}_\sim}}


\title{
         {\Large
                 {\bf
Exclusive $B \rar \rho \ell^+ \ell^-$ decay and 
Polarized lepton pair forward--backward asymmetries
                 }
         }
      }

\author{\vspace{1cm}\\
{\small T. M. Aliev \thanks
{e-mail: taliev@metu.edu.tr}\,\,,
V. Bashiry
\,\,,
M. Savc{\i} \thanks
{e-mail: savci@metu.edu.tr}} \\
{\small Physics Department, Middle East Technical University,
06531 Ankara, Turkey} }

\date{}

\begin{titlepage}
\maketitle
\thispagestyle{empty}

\begin{abstract}
The polarized lepton pair forward--backward asymmetries in $B \rar
\rho \ell^+ \ell^-$ decay using a general, model independent form
of the effective Hamiltonian is studied. The general expression for nine
double--polarization forward--backward asymmetries are calculated. 
The study of the forward--backward asymmetries of the doubly--polarized 
lepton pair proves to be very useful tool in looking for new physics 
beyond the standard model.
\end{abstract}

~~~PACS numbers: 13.20.He, 12.60.--i, 13.88.+e
\end{titlepage}

\section{Introduction}
Rare $B$ meson decays, induced by flavor changing neutral current (FCNC)
$b \rar s(d) \ell^+ \ell^-$ transitions provide a promising ground for
testing the gauge structure of the Standard Model (SM). These decays which 
are forbidden in the SM at tree level, occur at loop level and allow us
to check the prediction of the theory at quantum level. 
Moreover, these decays are also 
quite sensitive to the existence of new physics beyond the SM, since loops
with new particles can give contribution to these decays. The
new physics effects in rare decays can appear in two ways; one via
modification of the existing Wilson coefficients in the SM, or through the
introduction of some new operators with new coefficients. Theoretical
study of the $B \rar X_s \ell^+ \ell^-$ decays are relatively
more clean compared to their exclusive counterparts, since they are not
spoiled by nonperturbative long distance effects, while the corresponding 
exclusive channels are easier to measure experimentally.
Some of the most important exclusive FCNC decays are $B \rar K^\ast \gamma$
and $B \rar (\pi,\rho,K,K^\ast) \ell^+ \ell^-$ decays. The latter provides
potentially a very rich set of experimental observables, such as, lepton
pair forward--backward (FB) asymmetry, lepton polarizations, etc. 
Various kinematical distributions of such processes as
$B \rar K (K^\ast) \ell^+ \ell^-$ \cite{R6901,R6902,R6903}, 
$B \rar \pi (\rho) \ell^+ \ell^-$ \cite{R6904}, 
$B_{s,d} \rar \ell^+ \ell^-$ \cite{R6905} and 
$B_{s,d} \rar \gamma \ell^+ \ell^-$ \cite{R6906} have already been 
studied. Study of the forward--backward asymmetry, single polarization 
asymmetry, etc., which are experimentally measurable quantities, 
is an efficient approach that have been already been 
investigated in detail for the $B \rar K (K^\ast) \ell^+ \ell^-$ decay in
\cite{R6901,R6907,R6908,R6909,R6910,R6911,R6912} in fitting the parameters of
the SM and put constraints on new physics. It has
been pointed out in \cite{R6913} that some of the single lepton polarization
asymmetries might be quite small to be observed and might nor provide
sufficient number of observables in checking the structure of the effective
Hamiltonian. In \cite{R6910} the maximum number of independent observables
are constructed by taking both lepton polarizations into account
simultaneously. It is clear that, 
measurement of many more observables which would be useful in further
improvement of the parameters of the SM probing new physics beyond the SM.
It should be noted here that both lepton polarizations in the $B \rar K^\ast
\tau^+ \tau^-$ and $B \rar K \ell^+ \ell^-$ decays are studied in
\cite{R6914} and  \cite{R6915}, respectively.     
As has already been noted, one efficient way of establishing new physics
effects is studying forward--backward asymmetry in semileptonic $B \rar
K^\ast \ell^+ \ell^-$ decay, since, ${\cal A}_{FB}$ vanishes at specific
values of the dilepton invariant mass, and more essential than that,
this zero position of ${\cal A}_{FB}$ is known to be practically 
free of hadronic uncertainties \cite{R6912}. The decays of $B$ mesons
induced by the $b \rar d \ell^+ \ell^-$ transition are promising in looking
for CP violation since the CKM factors $V_{tb} V_{td}^\ast$, 
$V_{ub} V_{ud}^\ast$ and $V_{cb} V_{cd}^\ast$ in the SM are all of the same
order. For this reason CP violation is much more considerable in the decays
induced by $b \rar d$ transition. So, study of the exclusive decays $B_d \rar
(\pi,\rho,\eta) \ell^+ \ell^-$ are quite promising for the confirmation of
the CP violation and these decays have extensively been investigated in the
SM \cite{R6913} and beyond \cite{R6904}.   

The aim of the present work is studying the polarized forward--backward
asymmetry in the exclusive $B \rar \rho \ell^+ \ell^-$ decay using
a general form of the effective Hamiltonian, including all possible forms
of interactions. Here we would like to remind the reader that the influence
of new Wilson coefficients on various kinematical variables, such as branching
ratios, lepton pair forward--backward asymmetries and single lepton
polarization asymmetries for the inclusive $B \rar X_{s(d)} \ell^+ \ell^-$
decays (see first references in \cite{R6911,R6914,R6917}) and exclusive 
$B \rar K \ell^+ \ell^-,~\rho \ell^+ \ell^-,~\gamma  \ell^+ \ell^-,~
\pi  \ell^+ \ell^-,~\rho  \ell^+ \ell^-$
\cite{R6901,R6902,R6906,R6909,R6918,R6919} and pure leptonic 
$B \rar \ell^+ \ell^-$ decays \cite{R6905,R6920} have been studied comprehensively.

The paper is organized as follows. In section 2, using a general form
of the effective Hamiltonian, we obtain the matrix element
in terms of the form factors of the $B \rar \rho$ transition. In section 3
we derive the analytical results for the polarized forward--backward
asymmetry. Last section is devoted to the numerical analysis, discussion and
conclusions.

\section{Calculation of double lepton polarizations in 
$B \rar \rho \ell^+ \ell^-$ decay}

In this section we calculate the double lepton polarizations 
using a general form of the effective Hamiltonian. 
The $B \rar \rho \ell^+ \ell^-$ process is governed
by $b \rar d \ell^+ \ell^-$ transition at quark level. The matrix element
for the $b \rar d \ell^+ \ell^-$ can be written in terms of the
twelve model independent four--Fermi interactions in the following form:
\bea
\label{e6901}
{\cal H}_{eff} \es \frac{G_F\alpha}{\sqrt{2} \pi}
 V_{td}V_{tb}^\ast
\Bigg\{ C_{SL} \, \bar d i \sigma_{\mu\nu} \frac{q^\nu}{q^2}\, L \,b
\, \bar \ell \gamma^\mu \ell + C_{BR}\, \bar d i \sigma_{\mu\nu}
\frac{q^\nu}{q^2} \,R\, b \, \bar \ell \gamma^\mu \ell \nnb \\
\ar C_{LL}^{tot}\, \bar d_L \gamma_\mu b_L \,\bar \ell_L \gamma^\mu \ell_L +
C_{LR}^{tot} \,\bar d_L \gamma_\mu b_L \, \bar \ell_R \gamma^\mu \ell_R +
C_{RL} \,\bar d_R \gamma_\mu b_R \,\bar \ell_L \gamma^\mu \ell_L \nnb \\
\ar C_{RR} \,\bar d_R \gamma_\mu b_R \, \bar \ell_R \gamma^\mu \ell_R +
C_{LRLR} \, \bar d_L b_R \,\bar \ell_L \ell_R +
C_{RLLR} \,\bar d_R b_L \,\bar \ell_L \ell_R \nnb \\
\ar C_{LRRL} \,\bar d_L b_R \,\bar \ell_R \ell_L +
C_{RLRL} \,\bar d_R b_L \,\bar \ell_R \ell_L+
C_T\, \bar d \sigma_{\mu\nu} b \,\bar \ell \sigma^{\mu\nu}\ell \nnb \\
\ar i C_{TE}\,\epsilon^{\mu\nu\alpha\beta} \bar d \sigma_{\mu\nu} b \,
\bar \ell \sigma_{\alpha\beta} \ell  \Bigg\}~,
\eea
where $L$ and $R$ in (\ref{e6901}) are defined as
\bea
L = \frac{1-\gamma_5}{2} ~,~~~~~~ R = \frac{1+\gamma_5}{2}\nnb~,
\eea
and $C_X$ are the coefficients of the four--Fermi interactions.
The first two coefficients in Eq. (\ref{e6901}), $C_{SL}$ and $C_{BR}$, are 
the nonlocal Fermi interactions, which correspond to $-2 m_s C_7^{eff}$ and 
$-2 m_b C_7^{eff}$ in the SM, respectively. The following four terms
with coefficients $C_{LL}$, $C_{LR}$, $C_{RL}$ and $C_{RR}$ are the
vector type interactions. Two of these vector
interactions containing $C_{LL}^{tot}$ and $C_{LR}^{tot}$ do already
exist in the SM in the form $(C_9^{eff}-C_{10})$ and $(C_9^{eff}+C_{10})$.
Therefore, $C_{LL}^{tot}$ and $C_{LR}^{tot}$ can be written as
\bea
C_{LL}^{tot} &=& C_9^{eff} - C_{10} + C_{LL}~, \nnb \\
C_{LR}^{tot} &=& C_9^{eff} + C_{10} + C_{LR}~, \nnb
\eea
where $C_{LL}$ and $C_{LR}$ describe the contributions of the new physics.
The terms with
coefficients $C_{LRLR}$, $C_{RLLR}$, $C_{LRRL}$ and $C_{RLRL}$ describe
the scalar type interactions. The remaining last two terms lead by the
coefficients $C_T$ and $C_{TE}$, obviously, describe the tensor type
interactions.

It should be noted here that, in further analysis we will assume that all new
Wilson coefficients are real, as is the case in the SM, while only
$C_9^{eff}$ contains imaginary part and it is parametrized in the following
form
\bea
\label{e6902}
C_9^{eff} = \xi_1 + \lambda_u \xi_2~,
\eea
where
\bea
\lambda_u = \frac{V_{ub} V_{ud}^\ast}{V_{tb} V_{td}^\ast} \nnb~,
\eea
and
\bea
\label{e6903} 
\xi_1 \es 4.128  + 0.138 \omega(\hat{s}) + g(\hat{m}_c,\hat{s})
C_0(\hat{m}_b) 
- \frac{1}{2} g(\hat{m}_d,\hat{s}) (C_3 + C_4) \nnb \\
\ek \frac{1}{2}
g(\hat{m}_b,\hat{s}) (4 C_3 + 4 C_4 + 3C_5 + C_6)
+ \frac{2}{9} (3 C_3 + C_4 + 3C_5 + C_6)~,\nnb \\
\xi_2 \es [g(\hat{m}_c,\hat{s}) - g(\hat{m}_u,\hat{s})](3 C_1 + C_2)~,
\eea 
where $\hat{m}_q = m_q/m_b$, $\hat{s}=q^2$, $C_0(\mu)=3 C_1 + C_2 + 3 C_3 + 
C_4 + 3 C_5 + C_6$, and
\bea
\label{e6904}
\omega(\hat{s}) \es -\frac{2}{9} \pi^2 -\frac{4}{3} Li_2(\hat{s})-
\frac{2}{3} \ln (\hat{s}) \ln(1-\hat{s}) -
\frac{5+4\hat{s}}{3(1+2\hat{s})} \ln(1-\hat{s}) \nnb \\
\ek \frac{2 \hat{s}(1+\hat{s})(1-2\hat{s})}{3(1-\hat{s})^2(1+2\hat{s})}
\ln (\hat{s}) + \frac{5+9 \hat{s}-6 \hat{s}^2}{3(1-\hat{s})(1+2\hat{s})}~,
\eea
represents the $O(\alpha_s)$ correction coming from one gluon exchange
in the matrix element of the operator ${\cal O}_9$ \cite{R6921}, while the 
function $g(\hat{m}_q,\hat{s})$ represents one--loop corrections to the 
four--quark operators $O_1$--$O_6$ \cite{R6922}, whose form is
\bea
\label{e6905}
\lefteqn{
g(\hat{m}_q,\hat{s}) = -\frac{8}{9} \ln (\hat{m}_q) + \frac{8}{27} + 
\frac{4}{9} y_q - \frac{2}{9} (2+y_q)} \nnb \\
\ek \sqrt{\vel 1-y_q \ver} \Bigg\{ \theta(1-y_q)\Bigg[\ln \Bigg(\frac{1+\sqrt{1-y_q}}
{1-\sqrt{1-y_q}}\Bigg) -i\pi \Bigg] + \theta(y_q-1) 
\arctan \Bigg( \frac{1}{\sqrt{y_q-1}}\Bigg) \Bigg\}~,
\eea
where $y_q=4 \hat{m}_q^2/\hat{s}$.

In addition to the short distance contributions, $B\rar X_d \ell^+ \ell^-$
decay also receives long distance contributions, which have their origin in
the real $\bar{u}u$, $\bar{d}d$ and $\bar{c}c$ intermediate states, i.e.,
$\rho$, $\omega$ and $J/\psi$ family. There are four different approaches in
taking long distance contributions into consideration: a) HQET based
approach \cite{R6923}, b) AMM approach \cite{R6924}, c) LSW
approach \cite{R6925}, and d) KS approach \cite{R6926}. In the present work we
choose the AMM approach, in which these resonance contributions are
parametrized using the Breit--Wigner form for the resonant states. The
effective coefficient $C_9^{eff}$ including the $\rho$, $\omega$ and
$J/\psi$ resonances are defined as
\bea
\label{e6906}
C_9^{eff} = C_9(\mu) + Y_{res}(\hat{s})~,
\eea
where
\bea
\label{e6907}
Y_{res} \es -\frac{3\pi}{\alpha^2} \Bigg\{ \ga C^{(0)} (\mu) + \lambda_u
\left[3 C_1(\mu)+ C_2(\mu) \right] \dr 
\sum_{V_i=\psi} K_i \frac{\Gamma(V_i \rar \ell^+\ell^-) M_{V_i}}
{M_{V_i}^2-q^2-iM_{V_i}\Gamma_{V_i}} \nnb \\
\ek \lambda_u g(\hat{m}_u,\hat{s}) \left[3 C_1(\mu)+ C_2(\mu) \right] 
\sum_{V_i=\rho,\omega} \frac{\Gamma(V_i \rar \ell^+\ell^-) M_{V_i}} 
{M_{V_i}^2-q^2-iM_{V_i}\Gamma_{V_i}} \Bigg\}~.
\eea
The phenomenological factor $K_i$ has the universal value for the inclusive 
$B \rar X_{s(d)}\ell^+ \ell^-$ decay $K_i\simeq 2.3$ \cite{R6927}, which we use 
in our calculations.

The exclusive $B \rar \rho \ell^+ \ell^-$ decay is obtained from
the matrix elements of the quark operators in Eq. (\ref{e6901}) over meson
states, which can be parametrized in terms of the form factors.
Obviously, the following matrix elements  
\bea
&&\lla \rho\vel \bar d \gamma_\mu (1 \pm \gamma_5) 
b \ver B \rra~,\nnb \\
&&\lla \rho \vel \bar d i\sigma_{\mu\nu} q^\nu  
(1 \pm \gamma_5) b \ver B \rra~, \nnb \\
&&\lla \rho \vel \bar d (1 \pm \gamma_5) b 
\ver B \rra~, \nnb \\
&&\lla \rho \vel \bar d \sigma_{\mu\nu} b
\ver B \rra~, \nnb
\eea
are needed in obtaining the decay amplitude of the $B \rar \rho \ell^+ \ell^-$ decay. 
These matrix elements are defined as follows:
\bea
\lefteqn{
\label{e6908}
\lla \rho(p_{\rho},\varepsilon) \vel \bar d \gamma_\mu 
(1 \pm \gamma_5) b \ver B(p_B) \rra =} \nnb \\
&&- \epsilon_{\mu\nu\lambda\sigma} \varepsilon^{\ast\nu} p_{\rho}^\lambda q^\sigma
\frac{2 V(q^2)}{m_B+m_{\rho}} \pm i \varepsilon_\mu^\ast (m_B+m_{\rho})   
A_1(q^2) \\
&&\mp i (p_B + p_{\rho})_\mu (\varepsilon^\ast q)
\frac{A_2(q^2)}{m_B+m_{\rho}}
\mp i q_\mu \frac{2 m_{\rho}}{q^2} (\varepsilon^\ast q)
\left[A_3(q^2)-A_0(q^2)\right]~,  \nnb \\  \nnb \\
\lefteqn{
\label{e6909}
\lla \rho(p_{\rho},\varepsilon) \vel \bar d i \sigma_{\mu\nu} q^\nu
(1 \pm \gamma_5) b \ver B(p_B) \rra =} \nnb \\
&&4 \epsilon_{\mu\nu\lambda\sigma} 
\varepsilon^{\ast\nu} p_{\rho}^\lambda q^\sigma
T_1(q^2) \pm 2 i \left[ \varepsilon_\mu^\ast (m_B^2-m_{\rho}^2) -
(p_B + p_{\rho})_\mu (\varepsilon^\ast q) \right] T_2(q^2) \\
&&\pm 2 i (\varepsilon^\ast q) \left[ q_\mu -
(p_B + p_{\rho})_\mu \frac{q^2}{m_B^2-m_{\rho}^2} \right] 
T_3(q^2)~, \nnb \\  \nnb \\ 
\lefteqn{
\label{e6910}
\lla \rho(p_{\rho},\varepsilon) \vel \bar d \sigma_{\mu\nu} 
 b \ver B(p_B) \rra =} \nnb \\
&&i \epsilon_{\mu\nu\lambda\sigma}  \Bigg\{ - 2 T_1(q^2)
{\varepsilon^\ast}^\lambda (p_B + p_{\rho})^\sigma +
\frac{2}{q^2} (m_B^2-m_{\rho}^2) \Big[ T_1(q^2) - T_2(q^2) \Big] 
{\varepsilon^\ast}^\lambda q^\sigma \\
&&- \frac{4}{q^2} \Bigg[ T_1(q^2) - T_2(q^2) - \frac{q^2}{m_B^2-m_{\rho}^2} 
T_3(q^2) \Bigg] (\varepsilon^\ast q) p_{\rho}^\lambda q^\sigma \Bigg\}~. \nnb 
\eea
where $q = p_B-p_{\rho}$ is the momentum transfer and $\varepsilon$ is the
polarization vector of $\rho$ meson. 
In order to ensure finiteness of (\ref{e6908}) and (\ref{e6910}) at $q^2=0$, 
we assume that $A_3(q^2=0) = A_0(q^2=0)$ and $T_1(q^2=0) = T_2(q^2=0)$.
The matrix element $\lla \rho \vel \bar d (1 \pm \gamma_5 ) b \ver B \rra$
can be calculated by contracting both sides of Eq. (\ref{e6908}) 
with $q^\mu$ and using equation of motion. Neglecting the mass of the 
$d$ quark we get
\bea
\label{e6911}
\lla \rho(p_{\rho},\varepsilon) \vel \bar d (1 \pm \gamma_5) b \ver
B(p_B) \rra =
\frac{1}{m_b} \Big[ \mp 2i m_{\rho} (\varepsilon^\ast q)
A_0(q^2)\Big]~.
\eea
In deriving Eq. (\ref{e6911}) we have used the relationship
\bea
2 m_{\rho} A_3(q^2) = (m_B+m_{\rho}) A_1(q^2) -
(m_B-m_{\rho}) A_2(q^2)~, \nnb 
\eea
which follows from the equations of motion.

Using the definition of the form factors, as given above, the amplitude of
the  $B \rar \rho \ell^+ \ell^-$ decay can be written as 
\bea
\lefteqn{
\label{e6912}
{\cal M}(B\rightarrow \rho \ell^{+}\ell^{-}) =
\frac{G \alpha}{4 \sqrt{2} \pi} V_{tb} V_{td}^\ast }\nnb \\
&&\times \Bigg\{
\bar \ell \gamma^\mu(1-\gamma_5) \ell \, \Big[
-2 A_1 \epsilon_{\mu\nu\lambda\sigma} \varepsilon^{\ast\nu}
p_{\rho}^\lambda q^\sigma
 -i B_1 \varepsilon_\mu^\ast
+ i B_2 (\varepsilon^\ast q) (p_B+p_{\rho})_\mu
+ i B_3 (\varepsilon^\ast q) q_\mu  \Big] \nnb \\
&&+ \bar \ell \gamma^\mu(1+\gamma_5) \ell \, \Big[
-2 C_1 \epsilon_{\mu\nu\lambda\sigma} \varepsilon^{\ast\nu}
p_{\rho}^\lambda q^\sigma
 -i D_1 \varepsilon_\mu^\ast    
+ i D_2 (\varepsilon^\ast q) (p_B+p_{\rho})_\mu
+ i D_3 (\varepsilon^\ast q) q_\mu  \Big] \nnb \\
&&+\bar \ell (1-\gamma_5) \ell \Big[ i B_4 (\varepsilon^\ast
q)\Big]
+ \bar \ell (1+\gamma_5) \ell \Big[ i B_5 (\varepsilon^\ast
q)\Big]  \nnb \\
&&+4 \bar \ell \sigma^{\mu\nu}  \ell \Big( i C_T \epsilon_{\mu\nu\lambda\sigma}
\Big) \Big[ -2 T_1 {\varepsilon^\ast}^\lambda (p_B+p_{\rho})^\sigma +
B_6 {\varepsilon^\ast}^\lambda q^\sigma -
B_7 (\varepsilon^\ast q) {p_{\rho}}^\lambda q^\sigma \Big] \nnb \\
&&+16 C_{TE} \bar \ell \sigma_{\mu\nu}  \ell \Big[ -2 T_1
{\varepsilon^\ast}^\mu (p_B+p_{\rho})^\nu  +B_6 {\varepsilon^\ast}^\mu q^\nu -
B_7 (\varepsilon^\ast q) {p_{\rho}}^\mu q^\nu
\Bigg\}~,
\eea
where
\bea
\label{e6913}
A_1 &=& (C_{LL}^{tot} + C_{RL}) \frac{V}{m_B+m_{\rho}} -
2 (C_{BR}+C_{SL}) \frac{T_1}{q^2} ~, \nnb \\
B_1 &=& (C_{LL}^{tot} - C_{RL}) (m_B+m_{\rho}) A_1 - 2
(C_{BR}-C_{SL}) (m_B^2-m_{\rho}^2)
\frac{T_2}{q^2} ~, \nnb \\
B_2 &=& \frac{C_{LL}^{tot} - C_{RL}}{m_B+m_{\rho}} A_2 - 2
(C_{BR}-C_{SL})
\frac{1}{q^2}  \left[ T_2 + \frac{q^2}{m_B^2-m_{\rho}^2} T_3 \right]~,
\nnb \\
B_3 &=& 2 (C_{LL}^{tot} - C_{RL}) m_{\rho} \frac{A_3-A_0}{q^2}+
2 (C_{BR}-C_{SL}) \frac{T_3}{q^2} ~, \nnb \\
C_1 &=& A_1 ( C_{LL}^{tot} \rar C_{LR}^{tot}~,~~C_{RL} \rar
C_{RR})~,\nnb \\
D_1 &=& B_1 ( C_{LL}^{tot} \rar C_{LR}^{tot}~,~~C_{RL} \rar
C_{RR})~,\nnb \\
D_2 &=& B_2 ( C_{LL}^{tot} \rar C_{LR}^{tot}~,~~C_{RL} \rar
C_{RR})~,\nnb \\
D_3 &=& B_3 ( C_{LL}^{tot} \rar C_{LR}^{tot}~,~~C_{RL} \rar
C_{RR})~,\nnb \\
B_4 &=& - 2 ( C_{LRRL} - C_{RLRL}) \frac{ m_{\rho}}{m_b} A_0 ~,\nnb \\
B_5 &=& - 2 ( C_{LRLR} - C_{RLLR}) \frac{m_{\rho}}{m_b} A_0 ~,\nnb \\
B_6 &=& 2 (m_B^2-m_{\rho}^2) \frac{T_1-T_2}{q^2} ~,\nnb \\
B_7 &=& \frac{4}{q^2} \left( T_1-T_2 - 
\frac{q^2}{m_B^2-m_{\rho}^2} T_3 \right)~.   
\eea

From this expression of the decay amplitude, for the differential
decay width we get the following result:
\bea
\label{e6914}
\frac{d\Gamma}{d\hat{s}}(B \rar \rho \ell^+ \ell^-) =
\frac{G^2 \alpha^2 m_B}{2^{14} \pi^5}
\vel V_{tb}V_{td}^\ast \ver^2 \lambda^{1/2}(1,\hat{r},\hat{s}) v
\Delta(\hat{s})~,
\eea
with
\bea
\label{e6915}
\Delta \es
\frac{2}{3 \hat{r}_{\rho} \hat{s}} m_B^2
\,\mbox{\rm Re}\Big[
- 6 m_B \hat{m}_\ell \hat{s} \lambda
(B_1-D_1) (B_4^\ast - B_5^\ast) \nnb \\
\ek 12 m_B^2 \hat{m}_\ell^2 \hat{s} \lambda
\Big\{ B_4 B_5^\ast + (B_3-D_2-D_3) B_1^\ast - (B_2+B_3-D_3)
D_1^\ast \Big\} \nnb \\
\ar 6 m_B^3 \hat{m}_\ell \hat{s}
(1-\hat{r}_{\rho}) \lambda
(B_2-D_2) (B_4^\ast - B_5^\ast) \nnb \\
\ar 12 m_B^4 \hat{m}_\ell^2 \hat{s} 
(1-\hat{r}_{\rho}) \lambda
(B_2-D_2) (B_3^\ast-D_3^\ast) \nnb \\
\ar 6 m_B^3 \hat{m}_\ell \lambda \hat{s}^2
(B_4-B_5) (B_3^\ast-D_3^\ast) \nnb \\
\ar 48 \hat{m}_\ell^2 \hat{r}_{\rho} \hat{s} \Big\{ 3 B_1 D_1^\ast +
2 m_B^4 \lambda A_1 C_1^\ast \Big\} \nnb \\
\ar 48 m_B^5 \hat{m}_\ell \hat{s}\lambda^2
(B_2+D_2) B_7^\ast C_{TE}^\ast \nnb \\
\ek 16 m_B^4 \hat{r}_{\rho} \hat{s} (\hat{m}_\ell^2-\hat{s}) \lambda
\Big\{ \vel A_1\ver^2 + \vel C_1\ver^2 \Big\} \nnb \\
\ek m_B^2 \hat{s} (2 \hat{m}_\ell^2-\hat{s}) \lambda
\Big\{ \vel B_4\ver^2 + \vel B_5\ver^2 \Big\} \nnb \\
\ek 48 
m_B^3 \hat{m}_\ell \hat{s} (1-\hat{r}_{\rho}-\hat{s}) \lambda
\Big\{(B_1+D_1) B_7^\ast C_{TE}^\ast +
2 (B_2+D_2) B_6^\ast C_{TE}^\ast \Big\} \nnb \\
\ek 6 m_B^4 \hat{m}_\ell^2 \hat{s} \lambda
\Big\{ 2 (2+2\hat{r}_{\rho}-\hat{s}) B_2 D_2^\ast -
\hat{s} \vel (B_3-D_3)\ver^2 \Big\} \nnb \\
\ar 96
m_B \hat{m}_\ell \hat{s} (\lambda + 12 \hat{r}_{\rho} \hat{s})
(B_1+D_1) B_6^\ast C_{TE}^\ast\nnb \\
\ar 8 
m_B^2 \hat{s}^2 \Big\{
v^2 \vel C_T \ver^2 + 4 (3-2 v^2) \vel C_{TE} \ver^2 \Big\}
\Big\{4 (\lambda + 12 \hat{r}_{\rho} \hat{s}) \vel B_6 \ver^2 \nnb \\ 
\ek 4 m_B^2 \lambda (1-\hat{r}_{\rho}-\hat{s}) B_6 B_7^\ast
+ m_B^4 \lambda^2 \vel B_7 \ver^2  \Big\} \nnb \\
\ek 4 m_B^2 \lambda \Big\{
\hat{m}_\ell^2 (2 - 2 \hat{r}_{\rho} + \hat{s} ) +
\hat{s} (1 - \hat{r}_{\rho} - \hat{s} ) \Big\}
(B_1 B_2^\ast + D_1 D_2^\ast) \nnb \\
\ar \hat{s} \Big\{
6 \hat{r}_{\rho} \hat{s} (3+v^2) + \lambda (3-v^2)
\Big\} \Big\{ \vel B_1\ver^2 + \vel D_1\ver^2 \Big\} \nnb \\
\ek 2 m_B^4 \lambda \Big\{
\hat{m}_\ell^2 [\lambda - 3 (1-\hat{r}_{\rho})^2] - \lambda \hat{s} \Big\}
\Big\{ \vel B_2\ver^2 + \vel D_2\ver^2 \Big\} \nnb \\
\ar 128 m_B^2 \Big\{
4 \hat{m}_\ell^2 [ 20 \hat{r}_{\rho} \lambda - 
12 \hat{r}_{\rho} (1-\hat{r}_{\rho})^2 - \lambda \hat{s}] \nnb \\ 
\ar \hat{s} [ 4 \hat{r}_{\rho} \lambda + 12 \hat{r}_{\rho} (1-\hat{r}_{\rho})^2 +
\lambda \hat{s}] \Big\}
\vel C_T\ver^2 \vel T_1\ver^2 \nnb \\
\ar 512 m_B^2 \Big\{
\hat{s} [ 4 \hat{r}_{\rho} \lambda + 12 \hat{r}_{\rho} (1-\hat{r}_{\rho})^2 +
\lambda \hat{s}] \nnb \\ 
\ar 8 \hat{m}_\ell^2 [ 12 \hat{r}_{\rho} (1-\hat{r}_{\rho})^2 +
\lambda (\hat{s}-8 \hat{r}_{\rho})] \Big\}
\vel C_{TE}\ver^2 \vel T_1\ver^2 \nnb \\
\ek 64 m_B^2 \hat{s}^2 
\Big\{ v^2 \vel C_T \ver^2
+ 4 (3 - 2 v^2) \vel C_{TE} \ver^2 \Big\}
\Big\{ 2 [ \lambda  + 12 \hat{r}_{\rho} (1-\hat{r}_{\rho})]
B_6 T_1^\ast \nnb \\ 
\ek m_B^2 \lambda (1 + 3 \hat{r}_{\rho} - \hat{s}) 
B_7 T_1^\ast \Big\} \nnb \\
\ar 768  m_B^3 \hat{m}_\ell \hat{r}_{\rho} \hat{s}
\lambda (A_1 + C_1) C_T^\ast T_1^\ast \nnb \\
\ek 192 m_B \hat{m}_\ell \hat{s}
[ \lambda  + 12 \hat{r}_{\rho} (1-\hat{r}_{\rho})]
(B_1 + D_1) C_{TE}^\ast T_1^\ast \nnb \\
\ar 192 m_B^3 \hat{m}_\ell \hat{s} \lambda
(1 + 3 \hat{r}_{\rho} -\hat{s}) \lambda
(B_2 + D_2) C_{TE}^\ast T_1^\ast \Big]~,
\eea
where $\hat{s}=q^2/m_B^2$, $\hat{r}_{\rho}=m_{\rho}^2/m_B^2$ and
$\lambda(a,b,c)=a^2+b^2+c^2-2ab-2ac-2bc$,
$\hat{m}_\ell=m_\ell/m_B$, $v=\sqrt{1-4\hat{m}_\ell^2/\hat{s}}$ is the
final lepton velocity.

Using the matrix element for the $B \rar \rho \ell^+ \ell^-$ decay, our next
problem is to calculate the polarized $FB$ asymmetries. For this
purpose, we define the following orthogonal unit vectors $s_i^{\pm\mu}$ in
the rest frame of $\ell^\pm$, where $i=L,N$ or $T$ correspond to
longitudinal, normal, transversal polarization directions, respectively (see
also \cite{R6901,R6908,R6910,R6915}),
\bea
\label{e6916}   
s^{-\mu}_L \es \ga 0,\vec{e}_L^{\,-}\dr =
\ga 0,\frac{\vec{p}_-}{\vel\vec{p}_- \ver}\dr~, \nnb \\
s^{-\mu}_N \es \ga 0,\vec{e}_N^{\,-}\dr = \ga 0,\frac{\vec{p}_K\times
\vec{p}_-}{\vel \vec{p}_K\times \vec{p}_- \ver}\dr~, \nnb \\
s^{-\mu}_T \es \ga 0,\vec{e}_T^{\,-}\dr = \ga 0,\vec{e}_N^{\,-}
\times \vec{e}_L^{\,-} \dr~, \nnb \\
s^{+\mu}_L \es \ga 0,\vec{e}_L^{\,+}\dr =
\ga 0,\frac{\vec{p}_+}{\vel\vec{p}_+ \ver}\dr~, \nnb \\
s^{+\mu}_N \es \ga 0,\vec{e}_N^{\,+}\dr = \ga 0,\frac{\vec{p}_K\times
\vec{p}_+}{\vel \vec{p}_K\times \vec{p}_+ \ver}\dr~, \nnb \\
s^{+\mu}_T \es \ga 0,\vec{e}_T^{\,+}\dr = \ga 0,\vec{e}_N^{\,+}
\times \vec{e}_L^{\,+}\dr~,
\eea
where $\vec{p}_\mp$ and $\vec{p}_K$ are the three--momenta of the
leptons $\ell^\mp$ and $\rho$ meson in the
center of mass frame (CM) of $\ell^- \,\ell^+$ system, respectively.
Transformation of unit vectors from the rest frame of the leptons to CM
frame of leptons can be done by the Lorentz boost. Boosting of the
longitudinal unit vectors $s_L^{\pm\mu}$ yields
\bea
\label{e6917}
\ga s^{\mp\mu}_L \dr_{CM} \es \ga \frac{\vel\vec{p}_\mp \ver}{m_\ell}~,
\frac{E_\ell \vec{p}_\mp}{m_\ell \vel\vec{p}_\mp \ver}\dr~,
\eea
where $\vec{p}_+ = - \vec{p}_-$, $E_\ell$ and $m_\ell$ are the energy and mass
of leptons in the CM frame, respectively.
The remaining two unit vectors $s_N^{\pm\mu}$, $s_T^{\pm\mu}$ are unchanged
under Lorentz boost.

The definition of the unpolarized and normalized differential
forward--backward asymmetry is (see for example \cite{R6928})
\bea
\label{e6918}
{\cal A}_{FB} = \frac{\ds \int_{0}^{1} \frac{d^2\Gamma}{d\hat{s} dz} dz -
\int_{-1}^{0} \frac{d^2\Gamma}{d\hat{s} dz}dz }
{\ds \int_{0}^{1} \frac{d^2\Gamma}{d\hat{s} dz} dz +
\int_{-1}^{0} \frac{d^2\Gamma}{d\hat{s} dz} dz}~,
\eea
where $z=\cos\theta$ is the angle between $B$ meson and $\ell^-$ in the
center mass frame of leptons. When the spins of both leptons are taken into
account, the ${\cal A}_{FB}$ will be a function of the spins of the final
leptons and it is defined as
\bea
\label{e6919}
{\cal A}_{FB}^{ij}(\hat{s}) \es 
\Bigg(\frac{d\Gamma(\hat{s})}{d\hat{s}} \Bigg)^{-1}
\Bigg\{ \int_0^1 dz - \int_{-1}^0 dz \Bigg\}
\Bigg\{ 
\Bigg[
\frac{d^2\Gamma(\hat{s},\vec{s}^{\,-} = \vec{i},\vec{s}^{\,+} = \vec{j})}
{d\hat{s} dz} - 
\frac{d^2\Gamma(\hat{s},\vec{s}^{\,-} = \vec{i},\vec{s}^{\,+} = -\vec{j})} 
{d\hat{s} dz}
\Bigg] \nnb \\
\ek
\Bigg[
\frac{d^2\Gamma(\hat{s},\vec{s}^{\,-} = -\vec{i},\vec{s}^{\,+} = \vec{j})} 
{d\hat{s} dz} - 
\frac{d^2\Gamma(\hat{s},\vec{s}^{\,-} = -\vec{i},\vec{s}^{\,+} = -\vec{j})} 
{d\hat{s} dz}
\Bigg]
\Bigg\}~,\nnb \\ \nnb \\
\es 
{\cal A}_{FB}(\vec{s}^{\,-}=\vec{i},\vec{s}^{\,+}=\vec{j})   -
{\cal A}_{FB}(\vec{s}^{\,-}=\vec{i},\vec{s}^{\,+}=-\vec{j})  - 
{\cal A}_{FB}(\vec{s}^{\,-}=-\vec{i},\vec{s}^{\,+}=\vec{j})  \nnb \\
\ar   
{\cal A}_{FB}(\vec{s}^{\,-}=-\vec{i},\vec{s}^{\,+}=-\vec{j})~.   
\eea

Using these definitions for the double polarized $FB$ asymmetries, we get
the following results:   

\bea
\label{e6920}
{\cal A}_{FB}^{LL} \es 
\frac{2}{\hat{r}_{\rho}\Delta} m_B^3 \sqrt{\lambda} v \, \mbox{\rm Re}\Big[
- m_B^3 \hat{m}_\ell \lambda \Big\{ 4(B_1-D_1) B_7^\ast C_T^\ast -
(B_4+B_5) (B_2^\ast+D_2^\ast) \Big\} \nnb \\
\ar 4m_B^4 \hat{m}_\ell \lambda \Big\{
(1-\hat{r}_{\rho})(B_2-D_2) B_7^\ast C_T^\ast
+ \hat{s}(B_3-D_3) B_7^\ast C_T^\ast \Big\} \nnb \\
\ek \hat{m}_\ell (1-\hat{r}_{\rho} - \hat{s}) \Big\{     
B_1^\ast (B_4 + B_5 - 8 B_6 C_T) + D_1^\ast (B_4 + B_5 + 8 B_6 C_T)\Big\} \nnb \\
\ar 8 m_B \hat{r}_{\rho}\hat{s} (A_1 B_1^\ast - C_1 D_1^\ast)
+ 128 m_B^2 \hat{m}_\ell \hat{r}_{\rho} \hat{s}
(A_1 - C_1) B_6^\ast C_{TE}^\ast \nnb \\
\ar 2 m_B^3 \hat{s} \lambda \Big\{ (B_4 - B_5) B_7^\ast C_T^\ast +
2 (B_4 + B_5) B_7^\ast C_{TE}^\ast \Big\} \nnb \\
\ek 8 m_B^2 \hat{m}_\ell (1-\hat{r}_{\rho}) (1-\hat{r}_{\rho}-\hat{s})
(B_2 - D_2) B_6^\ast C_T^\ast \nnb \\
\ek 4 m_B (1-\hat{r}_{\rho}-\hat{s}) \hat{s} \Big\{
(B_4 - B_5) B_6^\ast C_T^\ast + 2 (B_4 + B_5) B_6^\ast C_{TE}^\ast \nnb \\ 
\ar 2 m_B \hat{m}_\ell (B_3 -D_3) B_6^\ast C_T^\ast 
\Big\} - 256 m_B^5 \hat{m}_\ell \hat{r}_{\rho} (1-\hat{r}_{\rho})
(A_1 - C_1) T_1^\ast C_{TE}^\ast \nnb \\
\ek 16 \hat{m}_\ell (1 - 5 \hat{r}_{\rho} - \hat{s})  
(B_1 - D_1) T_1^\ast C_T^\ast \nnb \\
\ar 16 m_B^2 \hat{m}_\ell (1 - \hat{r}_{\rho}) (1 + 3 \hat{r}_{\rho} - \hat{s}) 
(B_2 - D_2) T_1^\ast C_T^\ast \nnb \\
\ar 8 m_B  (1 + 3 \hat{r}_{\rho} - \hat{s}) \hat{s} \Big\{
2 (B_4 + B_5) T_1^\ast C_{TE}^\ast + (B_4 - B_5) T_1^\ast C_T^\ast \nnb \\
\ar 2 m_B \hat{m}_\ell (B_3 - D_3) T_1^\ast C_T^\ast \Big\}\Big]~,
\\ \nnb \\ \nnb
\label{e6921}
{\cal A}_{FB}^{LN} \es 
\frac{8}{3 \hat{r}_{\rho} \hat{s} \Delta} m_B^2 
\sqrt{\hat{s}}\lambda v \, \mbox{\rm Im}\Big[
- \hat{m}_\ell (B_1 D_1^\ast +
m_B^4 \lambda B_2 D_2^\ast )
+ 4 m_B^4 \hat{m}_\ell \hat{r}_{\rho} \sqrt{\hat{s}}
A_1 C_1^\ast \nnb \\
\ek 2 m_B \hat{s}
\Big\{ B_6 (C_T-2 C_{TE}) B_1^\ast +
B_6 (C_T+2 C_{TE}) D_1^\ast \Big\} \nnb \\
\ek m_B^5 \hat{s} \lambda 
\Big\{ B_7 (C_T-2 C_{TE}) B_2^\ast +
B_7 (C_T+2 C_{TE}) D_2^\ast \Big\} \nnb \\
\ek 16 m_B^2 \hat{m}_\ell \hat{s}
\Big( 4 \vel B_6 \ver^2 + m_B^4 \lambda \vel B_7 \ver^2 \Big)
C_T C_{TE}^\ast \nnb \\
\ar m_B^2 \hat{m}_\ell
(1 - \hat{r}_{\rho} - \hat{s})
(B_1 D_2^\ast + B_2 D_1^\ast) \nnb \\
\ar m_B^3
\hat{s} (1 - \hat{r}_{\rho} - \hat{s})
\Big\{ (B_1^\ast B_7 + 2 B_2^\ast B_6)(C_T - 2 C_{TE}) \nnb \\
\ar (D_1^\ast B_7 + 2 D_2^\ast B_6)(C_T + 
2 C_{TE}) \Big\} \nnb \\
\ek 64 m_B^2 \hat{m}_\ell \hat{s}
\Big\{ - m_B^2 (1 - \hat{r}_{\rho} - \hat{s}) \mbox{\rm Re}[B_6 B_7^\ast]
+ 4 \vel T_1 \ver^2 - 4 \mbox{\rm Re}[B_6 T_1^\ast] \nnb \\
\ar 2 m_B^2 (1 + 3 \hat{r}_{\rho} - \hat{s}) \mbox{\rm Re}[B_7 T_1^\ast]
\Big\} C_T C_{TE}^\ast \nnb \\
\ar 16 m_B^3 \hat{r}_{\rho} \hat{s}
\Big\{ (A_1 - C_1) C_T^\ast T_1^\ast -
2 (A_1 + C_1) C_{TE}^\ast T_1^\ast
\Big\} \nnb \\
\ar 4 m_B \hat{s}
\Big\{ B_1^\ast (C_T - 2 C_{TE}) T_1 +
D_1^\ast (C_T + 2 C_{TE}) T_1
\Big\} \nnb \\
\ek 4 m_B^3 \hat{s}
(1 + 3  \hat{r}_{\rho} - \hat{s})
\Big\{ B_2^\ast (C_T - 2 C_{TE}) T_1 +
D_2^\ast (C_T + 2 C_{TE}) T_1
\Big\} \Big]~, \\ \nnb \\ \nnb
\label{e6922}
{\cal A}_{FB}^{NL} \es
\frac{8}{3 \hat{r}_{\rho} \hat{s}\Delta} m_B^2 
\sqrt{\hat{s}}\lambda v \, \mbox{\rm Im}\Big[
- \hat{m}_\ell (B_1 D_1^\ast] +
m_B^4 \lambda B_2 D_2^\ast )
+ 4 m_B^2 \hat{m}_\ell \hat{r}_{\rho} \hat{s} 
A_1 C_1^\ast \nnb \\
\ar 2 m_B \hat{s}
\Big\{ B_6 (C_T+2 C_{TE}) B_1^\ast +
B_6 (C_T-2 C_{TE}) D_1^\ast \Big\} \nnb \\
\ar m_B^5 \hat{s} \lambda
\Big\{ B_7 (C_T+2 C_{TE}) B_2^\ast +
B_7 (C_T-2 C_{TE}) D_2^\ast \Big\}\nnb \\
\ar 16 m_B^2 \hat{m}_\ell \hat{s}
\Big( 4 \vel B_6 \ver^2 + m_B^4 \lambda \vel B_7 \ver^2 \Big)
C_T C_{TE}^\ast \nnb \\
\ar m_B^2 \hat{m}_\ell
(1 - \hat{r}_{\rho} - \hat{s})
(B_1 D_2^\ast + B_2 D_1^\ast) \nnb \\
\ek m_B^3
\hat{s} (1 - \hat{r}_{\rho} - \hat{s})
\Big\{ (B_1^\ast B_7 + 2 B_2^\ast B_6)(C_T + 2 C_{TE}) \nnb \\
\ar (D_1^\ast B_7 + 2 D_2^\ast B_6)(C_T - 2 C_{TE}) 
\Big\} \nnb \\
\ar 64 m_B^2 \hat{m}_\ell \hat{s}
\Big\{ - m_B^2 (1 - \hat{r}_{\rho} - \hat{s}) \mbox{\rm Re}[B_6 B_7^\ast]
+ 4 \vel T_1 \ver^2 - 4 \mbox{\rm Re}[B_6 T_1^\ast] \nnb \\
\ar 2 m_B^2 (1 + 3 \hat{r}_{\rho} - \hat{s}) \mbox{\rm Re}[B_7 T_1^\ast]
\Big\} C_T C_{TE}^\ast \nnb \\
\ar 16 m_B^3 \hat{r}_{\rho} \hat{s}
\Big\{ (A_1 - C_1) C_T^\ast T_1^\ast +
2 (A_1 + C_1) C_{TE}^\ast T_1^\ast
\Big\} \nnb \\
\ek 4 m_B \hat{s}
\Big\{ B_1^\ast (C_T + 2 C_{TE}) T_1 +
D_1^\ast (C_T - 2 C_{TE}) T_1
\Big\} \nnb \\
\ar 4m_B^3 \hat{s}
(1 + 3  \hat{r}_{\rho} - \hat{s})
\Big\{ B_2^\ast (C_T + 2 C_{TE}) T_1 +
D_2^\ast (C_T - 2 C_{TE}) T_1
\Big\} \Big]~, \\ \nnb \\ \nnb   
\label{e6923}
{\cal A}_{FB}^{LT} \es
\frac{4}{3 \hat{r}_{\rho} \hat{s}\Delta} m_B^2 
\sqrt{\hat{s}} \lambda \, \mbox{\rm Re} \Big[
- \hat{m}_\ell \Big\{ \vel B_1 + D_1 \ver^2 +
m_B^4 \lambda \vel B_2 + D_2 \ver^2 \Big\} \nnb \\
\ar 4  m_B^4 \hat{m}_\ell \hat{r}_{\rho} \hat{s}
\Big\{ \vel A_1 + C_1 \ver^2 \Big\} \nnb \\
\ek 64 m_B^2 \hat{m}_\ell \hat{s}
\vel C_{TE} \ver^2 \Big\{ 4 \vel B_6 \ver^2 + m_B^4 \lambda \vel B_7 \ver^2
- 4 m_B^2 (1 - \hat{r}_{\rho} - \hat{s}) B_6 B_7^\ast
\Big\} \nnb \\
\ar 2 m_B^2 \hat{m}_\ell
(1-\hat{r}_{\rho} - \hat{s})
(B_1+D_1) (B_2^\ast + D_2^\ast) \nnb \\
\ar 2 m_B^3 
(1-\hat{r}_{\rho} - \hat{s})
\Big\{ 4 \hat{m}_\ell^2 (2 B_2^\ast B_6 + B_1^\ast B_7)
(C_T + 2 C_{TE}) \nnb \\
\ek \hat{s} (2 B_2^\ast B_6 + B_1^\ast B_7) (C_T - 2 C_{TE})
\Big\} \nnb \\
\ek 4 m_B
\Big\{ 4 \hat{m}_\ell^2 \Big[ B_1^\ast B_6 (C_T + 2 C_{TE}) -
B_6 D_1^\ast (C_T - 2 C_{TE}) \Big] \nnb \\
\ek \hat{s} \Big[ B_1^\ast B_6 (C_T - 2 C_{TE}) -
B_6 D_1^\ast (C_T + 2 C_{TE}) \Big] \Big\} \nnb \\
\ek 2 m_B^5 \lambda
\Big\{ 4 \hat{m}_\ell^2 \Big[ B_2^\ast B_7 (C_T + 2 C_{TE}) -
B_7 D_2^\ast (C_T - 2 C_{TE}) \Big] \nnb \\
\ek \hat{s} \Big[ B_2^\ast B_7 (C_T - 2 C_{TE}) -
B_7 D_2^\ast (C_T + 2 C_{TE}) \Big] \Big\} \nnb \\
\ek 2 m_B^3       
(1-\hat{r}_{\rho} - \hat{s})
\Big\{ 4 \hat{m}_\ell^2
(2 B_6 D_2^\ast + B_7 D_1^\ast) (C_T - 2 C_{TE}) \nnb \\
\ek \hat{s} (2 B_6 D_2^\ast + B_7 D_1^\ast) (C_T + 2 C_{TE})
\Big\} \nnb \\
\ar 256 m_B^2 \hat{m}_\ell
\Big\{ 2 \hat{s} \vel C_{TE} \ver^2
\Big[ 2 B_6 T_1^\ast - 
m_B^2 (1+3 \hat{r}_{\rho} - \hat{s}) B_7 T_1^\ast 
\Big] \nnb \\ 
\ar 4 \vel T_1 \ver^2 \Big[ 
\hat{r}_{\rho} \vel C_T \ver^2 + 
(4 \hat{r}_{\rho} -\hat{s}) \vel C_{TE} \ver^2 \Big] \Big\} \nnb \\
\ar 32 m_B^3 \hat{r}_{\rho}
\Big\{ 4 \hat{m}_\ell^2
\Big[ (A_1+C_1) C_T^\ast T_1^\ast
+ 2 (A_1-C_1) C_{TE}^\ast T_1^\ast \Big] \nnb \\
\ar \hat{s} \Big[ A_1^\ast (C_T - 2 C_{TE}) T_1 +
C_1^\ast (C_T + 2 C_{TE}) T_1 \Big] 
\Big\} \nnb \\
\ar 8  m_B
\Big\{ 4 \hat{m}_\ell^2 (C_T + 2 C_{TE})-
\hat{s} (C_T - 2 C_{TE})\Big\} \Big\{B_1^\ast - m_B^2 (1+3 \hat{r}_{\rho} -
\hat{s}) B_2^\ast \Big\} T_1 \nnb \\
\ek 8 m_B
\Big\{ 4 \hat{m}_\ell^2 (C_T - 2 C_{TE})-
\hat{s} (C_T + 2 C_{TE})\Big\} \Big\{D_1^\ast - m_B^2 (1+3 \hat{r}_{\rho} -
\hat{s}) D_2^\ast \Big\} T_1 \Big]~,  \nnb \\ \\ \nnb
\label{e6924}
{\cal A}_{FB}^{TL} \es
\frac{4}{3 \hat{r}_{\rho} \hat{s}\Delta} m_B^2 
\sqrt{\hat{s}}\lambda \,\mbox{\rm Re}\Big[
\hat{m}_\ell \Big\{ \vel B_1 + D_1 \ver^2 +
m_B^4 \lambda \vel B_2 + D_2 \ver^2 \Big\} \nnb \\
\ek 4 m_B^4 \hat{m}_\ell \hat{r}_{\rho}
\Big\{ \vel A_1 + C_1 \ver^2 \Big\} \nnb \\
\ar 64 m_B^2 \hat{m}_\ell \hat{s}
\vel C_{TE} \ver^2 \Big\{ 4 \vel B_6 \ver^2 + m_B^4 \lambda \vel B_7 \ver^2
- 4 m_B^2 (1 - \hat{r}_{\rho} - \hat{s}) B_6 B_7^\ast
\Big\} \nnb \\
\ek 2 m_B^2 \hat{m}_\ell
(1-\hat{r}_{\rho} - \hat{s})
(B_1+D_1) (B_2^\ast + D_2^\ast) \nnb \\
\ar 2 m_B^3
(1-\hat{r}_{\rho} - \hat{s})
\Big\{ 4 \hat{m}_\ell^2 (2 B_2^\ast B_6 + B_1^\ast B_7)
(C_T - 2 C_{TE}) \nnb \\
\ek \hat{s} (2 B_2^\ast B_6 + B_1^\ast B_7) (C_T + 2 C_{TE})
\Big\} \nnb \\
\ek 4 m_B
\Big\{ 4 \hat{m}_\ell^2 \Big[ B_1^\ast B_6 (C_T - 2 C_{TE})
- B_6 D_1^\ast (C_T + 2 C_{TE}) \Big] \nnb \\
\ek \hat{s} \Big[ B_1^\ast B_6 (C_T + 2 C_{TE}) -
B_6 D_1^\ast (C_T - 2 C_{TE}) \Big] \Big\} \nnb \\
\ek 2m_B^5 \lambda
\Big\{ 4 \hat{m}_\ell^2 \Big[ B_2^\ast B_7 (C_T - 2 C_{TE}) -
B_7 D_2^\ast (C_T + 2 C_{TE}) \Big] \nnb \\
\ek \hat{s} \Big[ B_2^\ast B_7 (C_T + 2 C_{TE}) -
B_7 D_2^\ast (C_T - 2 C_{TE}) \Big] \Big\} \nnb \\
\ek 2 m_B^3      
(1-\hat{r}_{\rho} - \hat{s})
\Big\{ 4 \hat{m}_\ell^2
(2 B_6 D_2^\ast + B_7 D_1^\ast) (C_T + 2 C_{TE}) \nnb \\ 
\ek \hat{s} (2 B_6 D_2^\ast + B_7 D_1^\ast) (C_T - 2 C_{TE})
\Big\} \nnb \\
\ek 256 m_B^2 \hat{m}_\ell
\Big\{ 2 \hat{s} \vel C_{TE} \ver^2
\Big[ 2 B_6 T_1^\ast -
m_B^2 (1+3 \hat{r}_{\rho} - \hat{s}) B_7 T_1^\ast
\Big] \nnb \\ 
\ar 4 \vel T_1 \ver^2 \Big[ 
\hat{r}_{\rho} \vel C_T \ver^2 +
(4 \hat{r}_{\rho} -\hat{s}) \vel C_{TE} \ver^2 \Big] \Big\} \nnb \\
\ek 32 m_B^3 \hat{r}_{\rho}
\Big\{ 4 \hat{m}_\ell^2
\Big[ (A_1+C_1) C_T^\ast T_1^\ast 
- 2 (A_1-C_1) C_{TE}^\ast T_1^\ast \Big] \nnb \\
\ar \hat{s} \Big[ A_1^\ast (C_T + 2 C_{TE}) T_1 +
C_1^\ast (C_T - 2 C_{TE}) T_1 \Big] 
\Big\} \nnb \\
\ek 8 m_B
\Big\{ 4 \hat{m}_\ell^2 (C_T + 2 C_{TE})-
\hat{s} (C_T - 2 C_{TE})\Big\} \Big\{D_1^\ast - m_B^2 (1+3 \hat{r}_{\rho} -
\hat{s}) D_2^\ast \Big\} T_1 \nnb \\
\ar 8 m_B
\Big\{ 4 \hat{m}_\ell^2 (C_T - 2 C_{TE})-
\hat{s} (C_T + 2 C_{TE})\Big\} \Big\{B_1^\ast - m_B^2 (1+3 \hat{r}_{\rho} -
\hat{s}) B_2^\ast \Big\}~, T_1 \Big]  \nnb \\ \\ \nnb 
\label{e6925}
{\cal A}_{FB}^{NT} \es {\cal A}_{FB}^{TN} \nnb \\
\es \frac{2}{\hat{r}_{\rho} \hat{s}\Delta} m_B^2 \sqrt{\lambda}
\,\mbox{\rm Im}\Big[
m_B^3 \hat{m}_\ell \hat{s} \lambda
\Big\{ (B_4-B_5) (B_2^\ast+D_2^\ast) +
8 B_7 C_{TE} (B_1^\ast-D_1^\ast) \nnb \\ 
\ar 8 m_B^2 \hat{s} B_7^\ast C_{TE}^\ast (B_3-D_3)
\Big\} \nnb \\
\ek 2 m_B^4 \hat{m}_\ell^2 \hat{s} \lambda
(B_2+D_2) (B_3^\ast-D_3^\ast) \nnb \\
\ar 4 m_B^4 \hat{m}_\ell
(1-\hat{r}_{\rho}) \lambda
\Big\{ 2 m_B \hat{s} B_7^\ast C_{TE}^\ast (B_2-D_2) +
\hat{m}_\ell B_2 D_2^\ast \Big\} \nnb \\
\ar 2 m_B^2
\hat{m}_\ell^2 \hat{s} (1+ 3 \hat{r}_{\rho} - \hat{s})
(B_1 B_2^\ast - D_1 D_2^\ast) \nnb \\
\ar \hat{m}_\ell (1 - \hat{r}_{\rho} - \hat{s})
\Big\{
m_B \hat{s} \Big[ - B_1^\ast 
(B_4 - B_5 + 16 B_6 C_{TE}) \nnb \\
\ek D_1^\ast (B_4 - B_5 - 16 B_6 C_{TE})
+ 2 m_B \hat{m}_\ell (B_1+D_1) 
(B_3^\ast-D_3^\ast) \Big] \nnb \\
\ar 4 \Big[ \hat{m}_\ell B_1 D_1^\ast +
4 m_B^3 \hat{s}^2 B_6 C_{TE} (B_3^\ast-D_3^\ast) \Big]
\Big\} \nnb \\
\ek 16 m_B^3 \hat{m}_\ell \hat{s}
(1 - \hat{r}_{\rho}) (1 - \hat{r}_{\rho} - \hat{s})
(B_2 - D_2) B_6^\ast C_{TE}^\ast \nnb \\
\ar 2 m_B^2 \hat{m}_\ell^2
[\lambda +(1 - \hat{r}_{\rho}) (1 - \hat{r}_{\rho} - \hat{s})]
(B_1^\ast D_2 + B_2^\ast D_1) \nnb \\
\ar 32 m_B^3 \hat{m}_\ell \hat{s}
(1 - \hat{r}_{\rho}) (1 + 3 \hat{r}_{\rho} - \hat{s})
(B_2 -  D_2) C_{TE}^\ast T_1^\ast \nnb \\
\ek 8 m_B \hat{s}
(1 + 3 \hat{r}_{\rho} - \hat{s})
\Big\{ 4 \hat{m}_\ell (B_1-D_1) C_{TE}^\ast T_1^\ast
- 2 m_B \hat{s} (B_4-B_5) C_{TE}^\ast T_1^\ast \nnb \\
\ek 4 m_B^2 \hat{m}_\ell \hat{s} (B_3-D_3) C_{TE}^\ast T_1^\ast
+ m_B \hat{s} v^2 (B_4+B_5) C_T^\ast T_1^\ast
\Big\} \nnb \\
\ek 4 m_B^2  \hat{s}^2
(1 - \hat{r}_{\rho} - \hat{s})
\Big\{ 2 (B_4-B_5) B_6^\ast C_{TE}^\ast -
v^2 (B_4+B_5) B_6^\ast C_T^\ast
\Big\} \nnb \\
\ar 2 m_B^4 \hat{s}^2 \lambda
\Big\{ 2 (B_4-B_5) B_7^\ast C_{TE}^\ast -
v^2 (B_4+B_5) B_7^\ast C_T^\ast
\Big\} \Big]~, \\ \nnb \\ \nnb
\label{e6926}
{\cal A}_{FB}^{NN} \es - {\cal A}_{FB}^{TT} \nnb \\
\es \frac{2}{\hat{r}_{\rho}\Delta} m_B^3 \sqrt{\lambda} v
\,\mbox{\rm Re}\Big[
- m_B^2 \hat{m}_\ell \lambda
\Big\{ 4 (B_1-D_1) B_7^\ast C_T^\ast +
(B_2+D_2) (B_4^\ast + B_5^\ast) \Big\} \nnb \\
\ar 4 m_B^4 \hat{m}_\ell \lambda
\Big\{ (1-\hat{r}_{\rho}) (B_2-D_2) B_7^\ast C_T^\ast +
\hat{s} (B_3-D_3) B_7^\ast C_T^\ast \Big\} \nnb \\
\ar 2 m_B^3 \hat{s} \lambda
\Big\{ (B_4-B_5) B_7^\ast C_T^\ast -
2 (B_4+B_5) B_7^\ast C_{TE}^\ast
\Big\} \nnb \\
\ar \hat{m}_\ell
(1-\hat{r}_{\rho} - \hat{s})
\Big\{ B_1^\ast (B_4+B_5+8 B_6 C_T) \nnb \\
\ar D_1^\ast (B_4+B_5-8 B_6 C_T)
\Big\} \nnb \\
\ek 8 m_B^2 \hat{m}_\ell
(1-\hat{r}_{\rho}) (1-\hat{r}_{\rho} - \hat{s})
(B_2-D_2) B_6^\ast C_T^\ast \nnb \\
\ek 4 m_B \hat{s}
(1-\hat{r}_{\rho} - \hat{s})
\Big\{ (B_4-B_5) B_6^\ast C_T^\ast -
2 (B_4+B_5) B_6^\ast C_{TE}^\ast \nnb \\ 
\ar 2 m_B \hat{m}_\ell (B_3-D_3) B_6^\ast C_T^\ast
\Big\} \nnb \\
\ar 16 m_B^2 \hat{m}_\ell
(1-\hat{r}_{\rho}) (1+3 \hat{r}_{\rho} - \hat{s})
(B_2-D_2) C_T^\ast T_1^\ast \nnb \\
\ar 8 m_B \hat{s}
(1+3 \hat{r}_{\rho} - \hat{s})
\Big\{ (B_4-B_5) C_T^\ast T_1^\ast -
2 (B_4+B_5) C_{TE}^\ast T_1^\ast \Big\} \nnb \\
\ek 16\hat{m}_\ell
(1+3 \hat{r}_{\rho} - \hat{s})
(B_1-D_1) C_T^\ast T_1^\ast \nnb \\
\ar 16 m_B^2 \hat{m}_\ell \hat{s}
(1+3 \hat{r}_{\rho} - \hat{s})
(B_3-D_3) C_T^\ast T_1^\ast \Big]~.
\eea
In these expressions for ${\cal A}_{FB}^{ij}$, the first index in the superscript 
describes the polarization of lepton and the second index describes that of
anti--lepton.

\section{Numerical analysis}    

In this section we analyze the effects of the Wilson coefficients on
the polarized $FB$ asymmetry. The input parameters we use in our numerical
calculations are: $m_{\rho}=0.77~GeV$, $m_{\tau}=1.77~GeV$,
$m_{\mu}=0.106~GeV$, $m_{b}=4.8~GeV$, $m_{B}=5.26~GeV$ and 
$\Gamma_B = 4.22\times 10^{-13}~GeV$. 
For the values of the Wilson coefficients we use
$C_7^{SM}=-0.313,~C_9^{SM}=4.344$ and $C_{10}^{SM}=-4.669$. It should be
noted that the above--presented value for $C_9^{SM}$ corresponds only to
short distance contributions. In addition to the short distance
contributions, it receives long distance contributions which result from 
the conversion of $\bar{u}u$, $\bar{d}d$ and $\bar{c}c$ to the lepton pair. 
In order to minimize the hadronic uncertainties we will discard the regions
around low lying resonances $\rho$, $w$, $J/\psi$, $\psi^\prime$, 
$\psi^{\prime\prime}$, by dividing the $q^2$
region to low and high dilepton mass intervals:
\bea
\begin{array}{lc}
\mbox{\rm Region I:~~}&  1~GeV^2 \le q^2 \le 8~GeV^2~,\\
\mbox{\rm Region II:~~}& 14.5~GeV^2 \le q^2 \le (m_B-m_\rho)^2~,
\end{array} \nnb
\eea
where the contributions of the higher $\psi$ resonances do still exist in
the second region.
For the form factors we have used the light cone QCD sum rules results
\cite{R6924,R6929}. As a result of the analysis carried out in this scheme,
the $q^2$ dependence of the form factors can be represented in terms of
three parameters as
\bea
F(q^2) = \frac{F(0)}{\ds 1-a_F \hat{s} + b_F 
    \hat{s}^2}~, \nnb
\eea
where the values of parameters $F(0)$, $a_F$ and $b_F$ for the
$B \rar \rho$ decay are listed in Table 1. 

\begin{table}[h]
\renewcommand{\arraystretch}{1.5}
\addtolength{\arraycolsep}{3pt}
$$
\begin{array}{|l|ccc|}
\hline
& F(0) & a_F & b_F \\ \hline
A_0^{B \rar \rho} &
\phantom{-}0.372 \pm 0.04 & 1.40 & \phantom{-}0.437 \\
A_1^{B \rar \rho} &
\phantom{-}0.261 \pm 0.04 & 0.29 & -0.415 \\
A_2^{B \rar \rho} &
\phantom{-}0.223 \pm 0.03 & 0.93 & -0.092\\
V^{B \rar \rho} &
 \phantom{-}0.338 \pm 0.05 & 1.37 & \phantom{-}0.315\\
T_1^{B \rar \rho} &
  \phantom{-}0.285 \pm 0.04 & 1.41 & \phantom{-}0.361\\
T_2^{B \rar \rho} &
 \phantom{-}0.285 \pm 0.04 & 0.28 & -0.500\\
T_3^{B \rar \rho} &
 \phantom{-}0.202 \pm 0.04 & 1.06 & -0.076\\ \hline
\end{array}
$$       
\caption{$B$ meson decay form factors in a three-parameter fit, where the
radiative corrections to the leading twist contribution and SU(3) breaking
effects are taken into account.}
\renewcommand{\arraystretch}{1}
\addtolength{\arraycolsep}{-3pt}
\end{table}

In further numerical analysis,   
the values of the new Wilson coefficients are needed, and in the present
analysis we will vary them in the range $-\ve C_{10}\ve \le \ve C_i\ve
\le \ve C_{10}\ve$. The experimental value of the branching ratio of the $B
\rar K^\ast \ell^+ \ell^-$ decay \cite{R6930,R6931} and the bound on the
branching ratio of the $B \rar \mu^+ \mu^-$ \cite{R6932} suggest that this 
is the right order of magnitude for the vector and scalar interaction 
coefficients. It should be noted here that the experimental results lead to 
stronger restrictions on some of the Wilson coefficients, namely $-1.5 \le
C_T \le 1.5$, $-3.3 \le C_{TE} \le 2.6$, $-2 \le C_{LL}$, $C_{RL} \le 2.3$,
while the remaining coefficients vary in the range $-4 \le C_X \le 4$. 

As is obvious from the explicit expressions of the forward--backward
asymmetries, they depend both on $q^2$ and the new Wilson coefficients
$C_X$. As a result of this, it might be difficult to study the dependence of
the polarized forward--backward asymmetries ${\cal A}_{FB}^{ij}$ on these
parameters simultaneously. Therefore, it is necessary to eliminate the
dependence of ${\cal A}_{FB}^{ij}$ on one of the parameters.
We eliminate the dependence of the polarized 
${\cal A}_{FB}^{ij}$ on $q^2$ by performing integration over $q^2$ in the
kinematically allowed region, so that the polarized forward--backward
asymmetry is said to be averaged. The averaged polarized forward--backward
asymmetry is defined as
\bea
\lla {\cal A}_{FB}^{ij} \rra = \frac{\ds
\int_{q_{min}^2}^{q_{max}^2} {\cal A}_{FB}^{ij} 
\frac{d {\cal B}}{dq^2} dq^2 }
{\ds\int_{q_{min}^2}^{q_{max}^2} \frac{d {\cal B}}{dq^2} 
dq^2 }~.\nnb
\eea 

In Figs. (1) and (2), we present the dependence of $\lla {\cal A}_{FB}^{LL} \rra$
on $C_X$ for the $B\rar \rho \mu^+ \mu^-$ decay, in the Regions I and II, 
respectively. The common intersection 
point of all curves corresponds to the SM case. We observe from this figure
that, $\lla {\cal A}_{FB}^{LL} \rra$ has practically symmetric behavior in
regard to its 
dependence on $C_T$ and $C_{TE}$ with respect to zero position. We see from
Fig. (1) that, $\lla {\cal A}_{FB}^{LL} \rra$ for the $B\rar \rho \mu^+
\mu^-$ decay is very strongly dependent on $C_{LL}$, $C_{LR}$, $C_{T}$,
$C_{TE}$ and on $C_{RL}$ when $C_{RL}>1$. There are certain regions of $C_X$
where the magnitude of $\lla {\cal A}_{FB}^{LL} \rra$ is, more or less, two
times larger than compared to its value in the SM. This fact is a direct
indication of the confirmation of new physics beyond the SM which can be
attributed to the existence of new vector type interaction.
$\lla {\cal A}_{FB}^{LL} \rra$ behaves in the same way in region II 
as it does in Region I, except for the scalar interactions and vector 
interaction with coefficient $C_{RR}$ (see Fig. (2)), and being quite
sensitive to the rest of the remaining new Wilson coefficients. It is
interesting to observe that the sign of $\lla {\cal A}_{FB}^{LL} \rra$ is
negative (positive) in Region I (Region II) for all values of $C_X$.
  
Figs. (3) and (4) depict the dependence of $\lla {\cal A}_{FB}^{LT} \rra$
on $C_X$ for the $B\rar \rho \mu^+ \mu^-$ decay, in Regions I and II,
respectively. We observe from Fig. (3) that, except scalar interactions,
$\lla {\cal A}_{FB}^{LT} \rra$ is quite sensitive to the existence of the
remaining ones. More important than that is the presence of regions of the
new Wilson coefficients where $\lla {\cal A}_{FB}^{LT} \rra$ changes its
sign, while in the SM case the sign of the $\lla {\cal A}_{FB}^{LT} \rra$
is never switched. So, study of the magnitude and sign of $\lla {\cal
A}_{FB}^{LT} \rra$ can serve as a good test for looking new physics beyond
the SM. In region II, $\lla {\cal A}_{FB}^{LT} \rra$ is strongly dependent
only on tensor interactions (see Fig. (4)).

The dependence of $\lla {\cal A}_{FB}^{TL} \rra$ on $C_X$ for the $B\rar
\rho \mu^+ \mu^-$ decay, is given in Fig. (5) in Region I, and Fig. (6) in
Region II, respectively. We observe from these figures that $\lla {\cal
A}_{FB}^{TL} \rra$ exhibits strong dependence only on tensor interactions,
and especially there is a region of $C_{TE}$ where $\lla {\cal A}_{FB}^{TL}
\rra$ exceeds the SM prediction more than one order of magnitude. Moreover,
when $C_{T}$ and $C_{TE}$ is negative (positive) the sign of 
$\lla {\cal A}_{FB}^{TL} \rra$ is positive (negative). Hence,
determination of its magnitude and sign is an unambiguous confirmation of
the existence of tensor interaction. Similarly, Fig. (6) depicts strong
dependence of $\lla {\cal A}_{FB}^{TL} \rra$ on tensor interactions. When
$C_{TE}$ is negative the sign of  $\lla {\cal A}_{FB}^{TL} \rra$ is
positive, and when $C_{TE}$ is positive it is negative.
Analogous behavior is observed for $C_{T}$, except for quite a narrow 
region.   

All remaining forward--backward asymmetries for the $B\rar \rho \mu^+ \mu^-$
decay are numerically very small and for this reason we do not present them.

The study of the dependence of forward--backward asymmetry for the 
$B\rar \rho \tau^+ \tau^-$ decay gives richer information. In Figs. (7),
(8), (9) and (10) we present the dependence of $\lla {\cal A}_{FB}^{LL}
\rra$, $\lla {\cal A}_{FB}^{LT} \rra$, $\lla {\cal A}_{FB}^{TL} \rra$,
$\lla {\cal A}_{FB}^{NT} \rra = \lla {\cal A}_{FB}^{TN} \rra$ and
$\lla {\cal A}_{FB}^{NN} \rra = -\lla {\cal A}_{FB}^{TT} \rra$ on new Wilson
coefficients, respectively. We see from Fig. (7) that when $C_X$ is
negative, $\lla {\cal A}^{LL}\rra > \lla {\cal A}_{SM}^{LL} \rra$ only for 
the scalar interaction $C_{RLLR}$. Also, when $C_X$ is positive,
$\lla {\cal A}^{LL}\rra > \lla {\cal A}_{SM}^{LL} \rra$ for the coefficients
$C_{RL}$, $C_{LRLR}$. Fig. (8) depicts that $\lla {\cal A}_{FB}^{LT} \rra$
is strongly dependent on tensor interactions, as well as on vector
interactions with the coefficients $C_{RL}$ and $C_{RR}$ when they get
positive values. We observe from Fig. (9) that, $\lla {\cal A}_{FB}^{NT}
\rra = \lla {\cal A}_{FB}^{TN}\rra$ both exhibit strong dependence on all new
Wilson coefficients. As can easily be observed from Fig. (10), 
$\lla {\cal A}_{FB}^{NN} \rra = -\lla {\cal A}_{FB}^{TT} \rra$ both are very
sensitive to the presence of tensor and scalar interactions.

It follows from these results that few of the polarized forward--backward
asymmetries show considerable departure from the SM predictions and these 
ones are strongly dependent on different types of interactions. Hence, the
study of these quantities can play crucial role in establishing new physics
beyond the SM.       
 
At the end of this section, we would like to discuss the following problem.
It is clear that the existence of new physics can more easily be checked
through branching ratio measurements. In this connection there follows the
question: could there be a situation in which the branching ratio coincides
with that of the SM result, while polarized forward--backward asymmetry does
not? In order to answer this question we study the correlation between the
$\lla {\cal A}_{FB}^{ij} \rra$ and the branching ratio ${\cal B}$. We can
briefly summarize the results of our numerical analysis as follows:
As far as $B\rar \rho \mu^+ \mu^-$ decay is concerned, except for a very 
narrow region of $C_{RR}$, such a region is absent
for all new Wilson coefficients for the asymmetries
$\lla {\cal A}_{FB}^{LL} \rra$, $\lla {\cal A}_{FB}^{LT} \rra$ and $\lla
{\cal A}_{FB}^{TL} \rra$.

The $B\rar \rho \tau^+ \tau^-$ decay is more informative for this aim. In
Figs. (11) and (12) we present the dependence of $\lla {\cal A}_{FB}^{LL}
\rra$ and $\lla {\cal A}_{FB}^{LT} \rra$ on the branching ratio. It follows
from these figures that, there indeed exists certain regions of $C_X$ for
which the polarized forward--backward asymmetry differs from the SM
prediction, while the branching ratio coincides with that of the SM result.
We also note that, such a region exists for the asymmetries
$\lla {\cal A}_{FB}^{NN} \rra = -\lla {\cal A}_{FB}^{TT} \rra$ as well, 
only for the tensor interaction. 
 
In conclusion, in this work we investigate the forward--backward
asymmetries when both leptons are polarized, using a general, model
independent form of the effective Hamiltonian. We see that the
study of the zero position of $\lla {\cal A}_{FB}^{LL} \rra$ can give
unambiguous conformation of the new physics beyond the SM, since when new
physics effects are taken into account, the results are shifted with respect to 
their zero positions in the SM. We also find out that the polarized 
${\cal A}_{FB}$ is quite sensitive to the existence of some of the new
Wilson coefficients. We see that there exist certain regions of some of the new
Wilson coefficients for which, only study of the polarized forward--backward
asymmetry gives invaluable information in establishing new physics beyond
the SM.

\newpage

\section*{Figure captions}
{\bf Fig. (1)} The dependence of the averaged forward--backward
double--lepton polarization asymmetry $\lla {\cal A}_{FB}^{LL} \rra$
on the new Wilson coefficients $C_X$, for the $B \rar \rho \mu^+ \mu^-$
decay, in Region I.\\ \\
{\bf Fig. (2)} The same as in Fig. (1), but in Region II.\\ \\
{\bf Fig. (3)} The same as in Fig. (1), but for the averaged
forward--backward double--lepton polarization asymmetry
$\lla {\cal A}_{FB}^{LL} \rra$.\\ \\
{\bf Fig. (4)} The same as in Fig. (3), but in Region II.\\ \\
{\bf Fig. (5)} The same as in Fig. (1), but for the averaged 
forward--backward double--lepton polarization asymmetry
$\lla {\cal A}_{FB}^{TL} \rra$.\\ \\
{\bf Fig. (6)} The same as in Fig. (5), but in Region II.\\ \\
{\bf Fig. (7)} The dependence of the averaged forward--backward
double--lepton polarization asymmetry $\lla {\cal A}_{FB}^{LL} \rra$
on the new Wilson coefficients $C_X$, for the $B \rar \rho \tau^+ \tau^-$
decay, in Region II.\\ \\
{\bf Fig. (8)} The same as in Fig. (7), but for the averaged
forward--backward double--lepton polarization asymmetry
$\lla {\cal A}_{FB}^{LT} \rra$.\\ \\
{\bf Fig. (9)} The same as in Fig. (7), but for the averaged
forward--backward double--lepton polarization asymmetry
$\lla {\cal A}_{FB}^{NT} \rra = \lla {\cal A}_{FB}^{TN} \rra$.\\ \\
{\bf Fig. (10)} The same as in Fig. (7), but for the averaged
forward--backward double--lepton polarization asymmetry
$\lla {\cal A}_{FB}^{NN} \rra = - \lla {\cal A}_{FB}^{TT} \rra$.\\ \\
{\bf Fig. (11)} Parametric plot of the correlation between the averaged 
forward--backward double--lepton polarization asymmetry
$\lla {\cal A}_{FB}^{LL} \rra$ and the branching ratio for the
$B \rar \rho \tau^+ \tau^-$ decay, in Region II.\\ \\
{\bf Fig. (12)} Parametric plot of the correlation between the averaged
forward--backward double--lepton polarization asymmetry
$\lla {\cal A}_{FB}^{LT} \rra$ and the branching ratio for the
$B \rar \rho \tau^+ \tau^-$ decay, in Region II.

\newpage

\begin{figure}
\vskip 1.5 cm
    \includegraphics{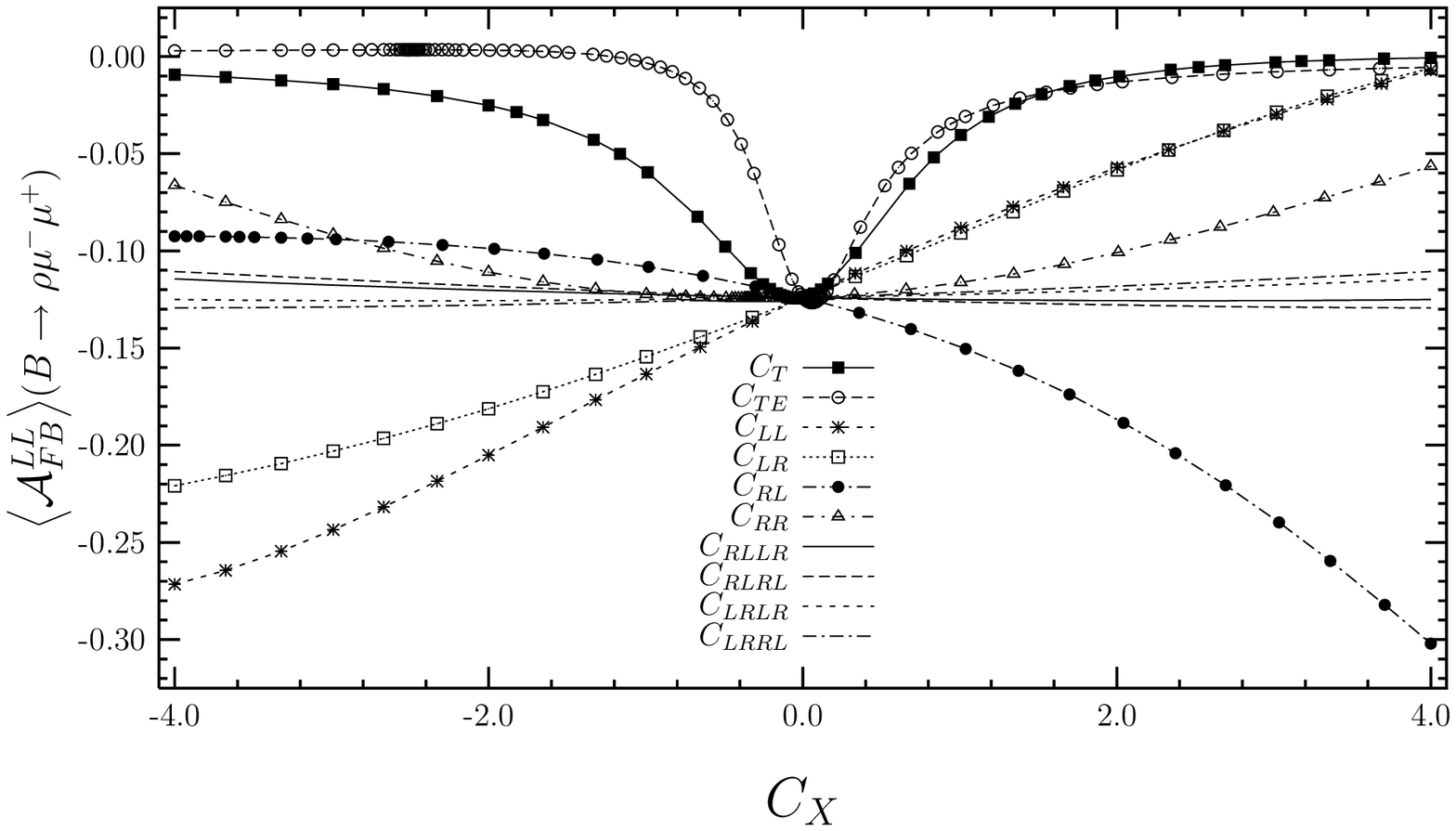}
\vskip 7.8cm
\caption{}
\end{figure}

\begin{figure}
\vskip 2.5 cm
    \includegraphics{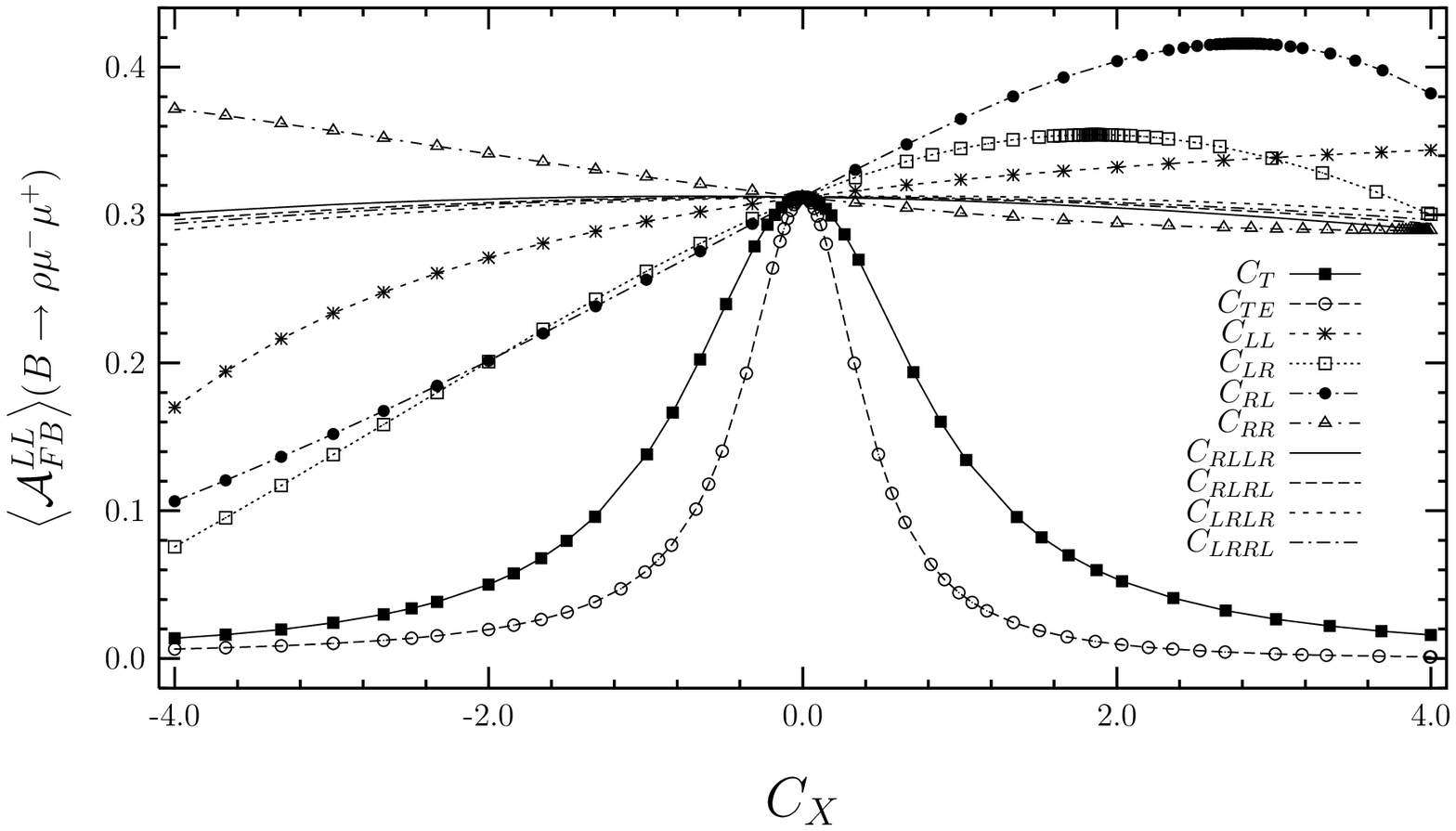}
\vskip 7.8 cm
\caption{}
\end{figure}

\begin{figure}
\vskip 1.5 cm
    \includegraphics{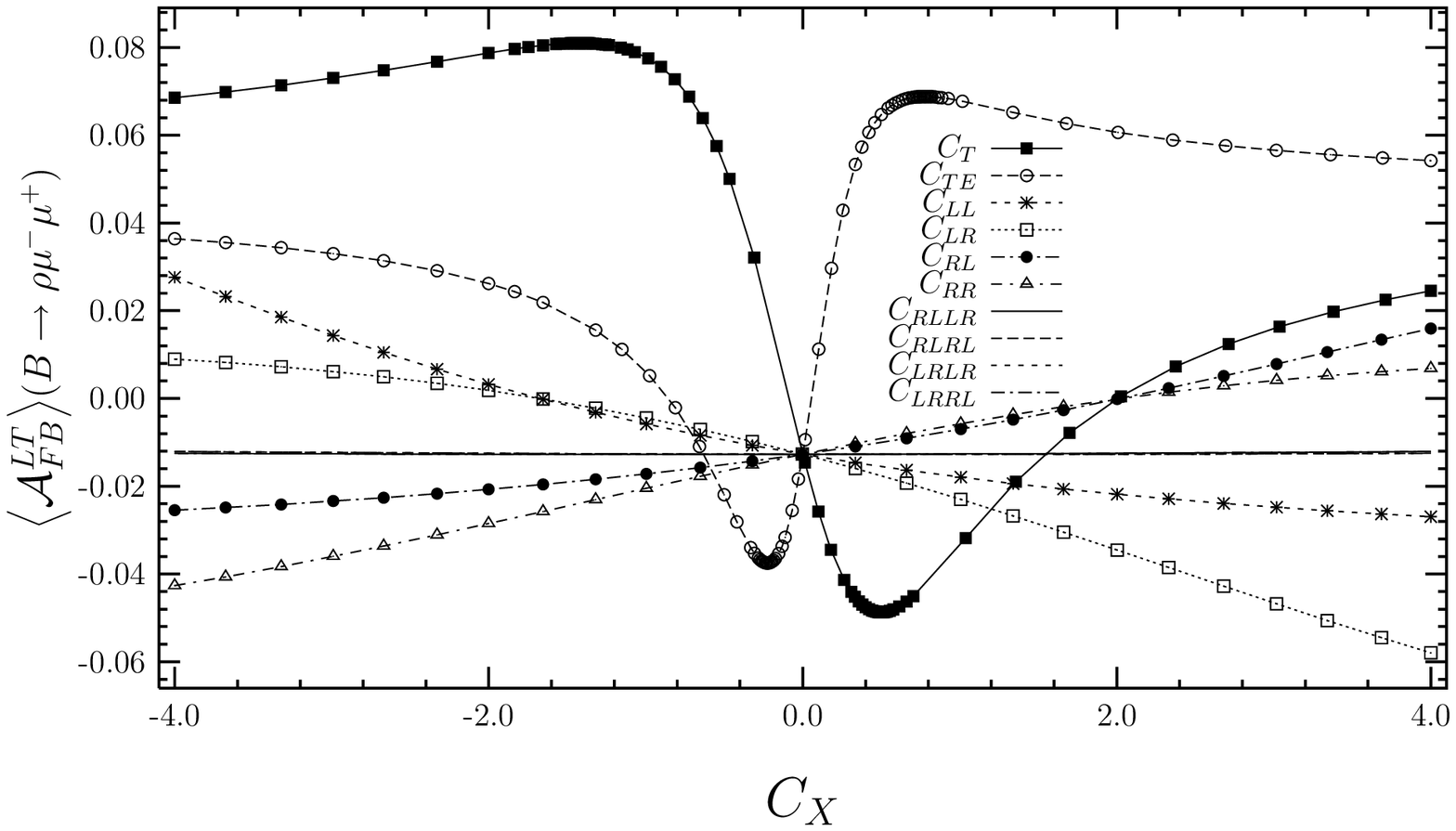}
\vskip 7.8cm
\caption{}
\end{figure}

\begin{figure}
\vskip 2.5 cm
    \includegraphics{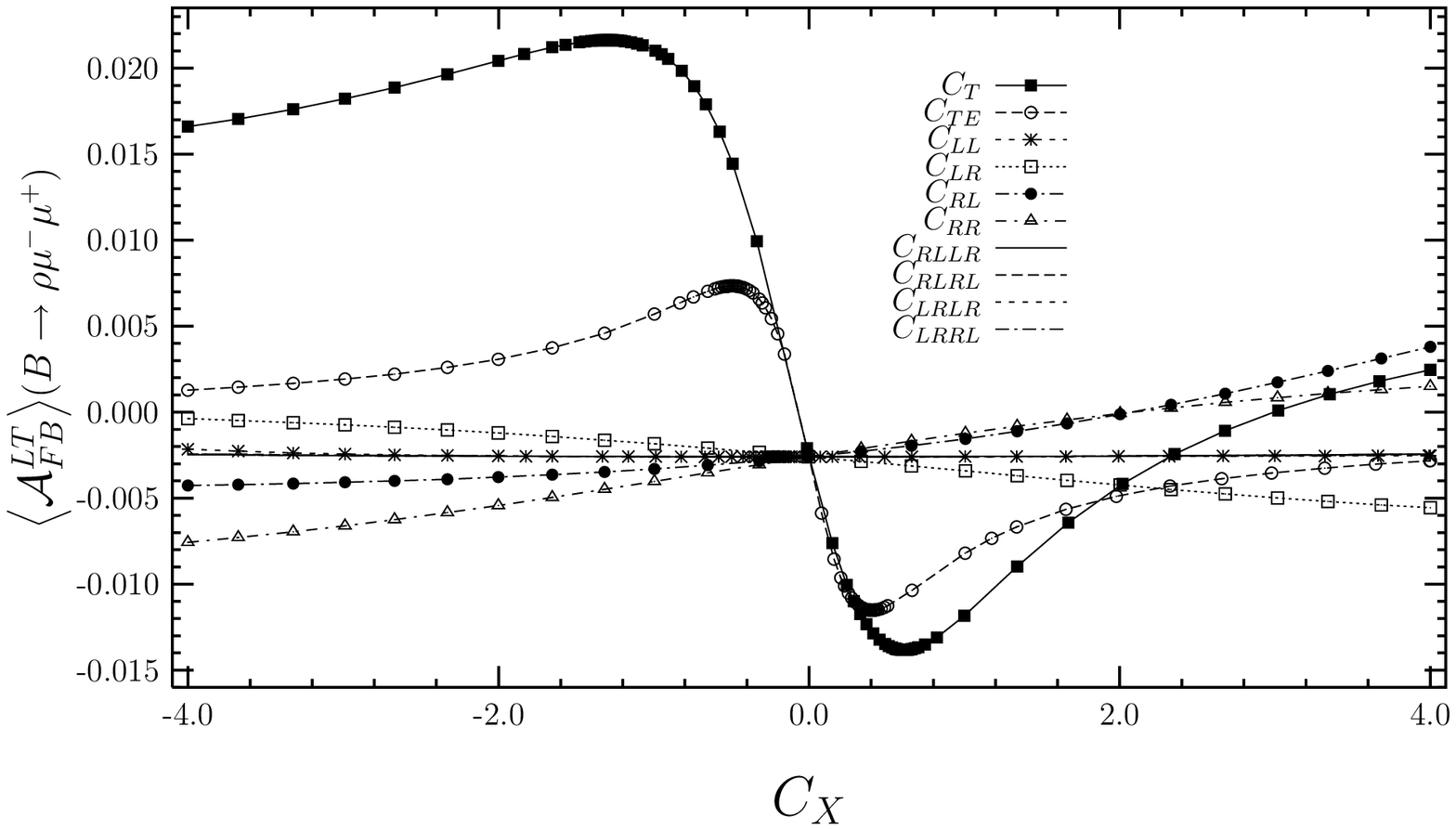}
\vskip 7.8 cm
\caption{}
\end{figure}

\begin{figure}
\vskip 2.5 cm
    \includegraphics{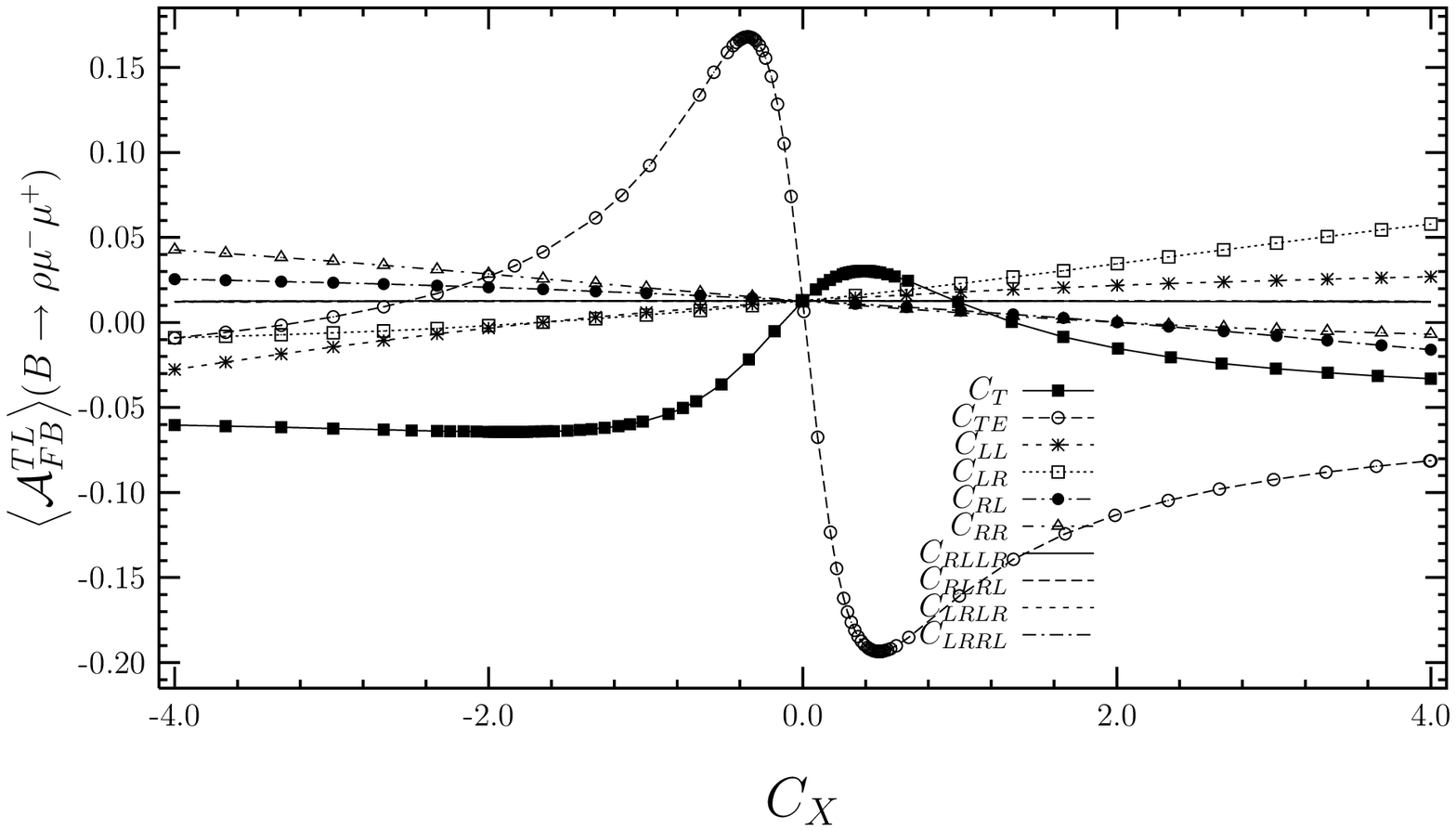}
\vskip 7.8 cm
\caption{}
\end{figure}

\begin{figure}
\vskip 1.5 cm
    \includegraphics{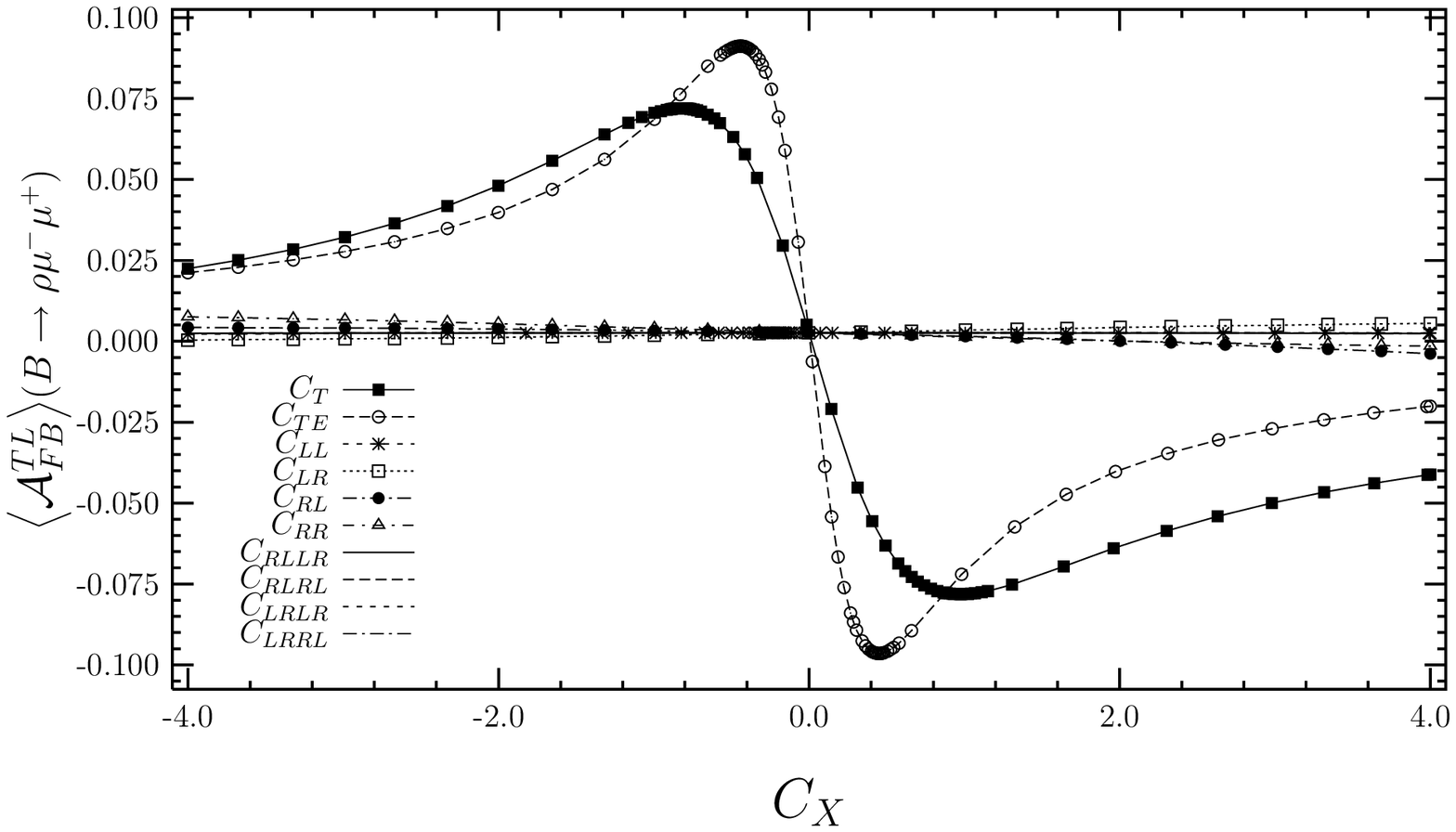}
\vskip 7.8cm
\caption{}
\end{figure}

\begin{figure}
\vskip 2.5 cm
    \includegraphics{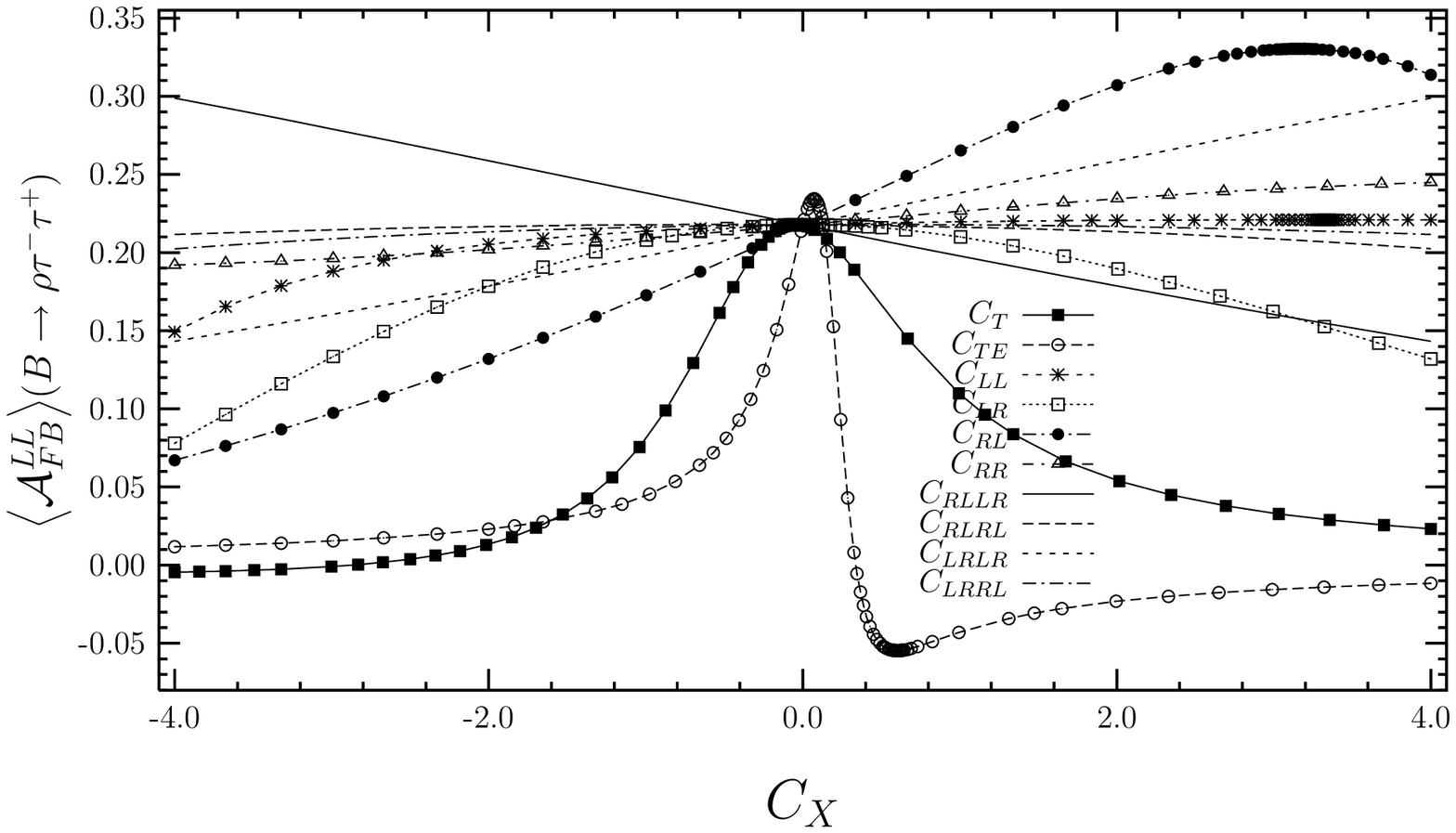}
\vskip 7.8 cm
\caption{}
\end{figure}

\begin{figure}
\vskip 1.5 cm
    \includegraphics{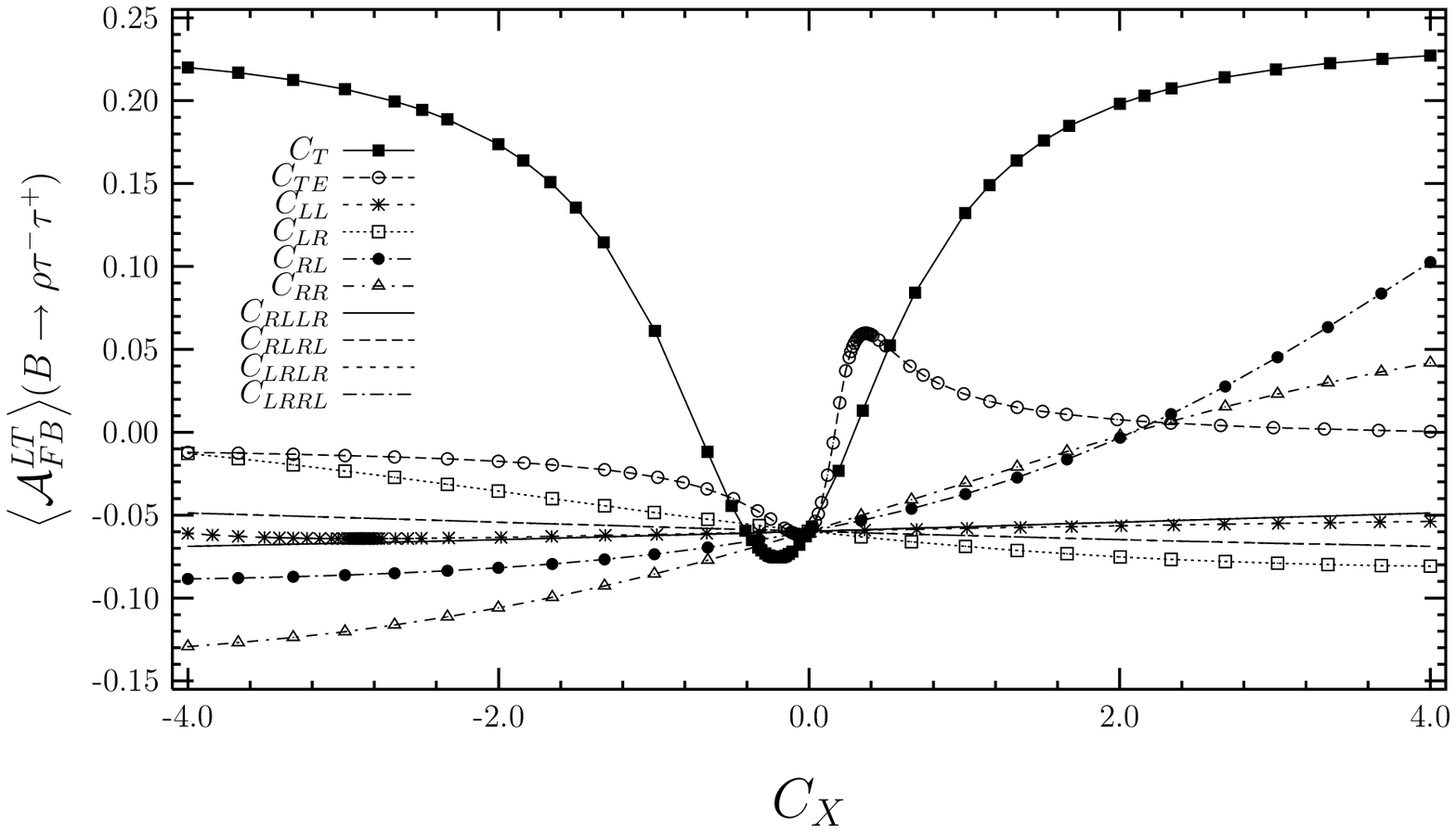}
\vskip 7.8cm
\caption{}
\end{figure}

\begin{figure}
\vskip 2.5 cm
    \includegraphics{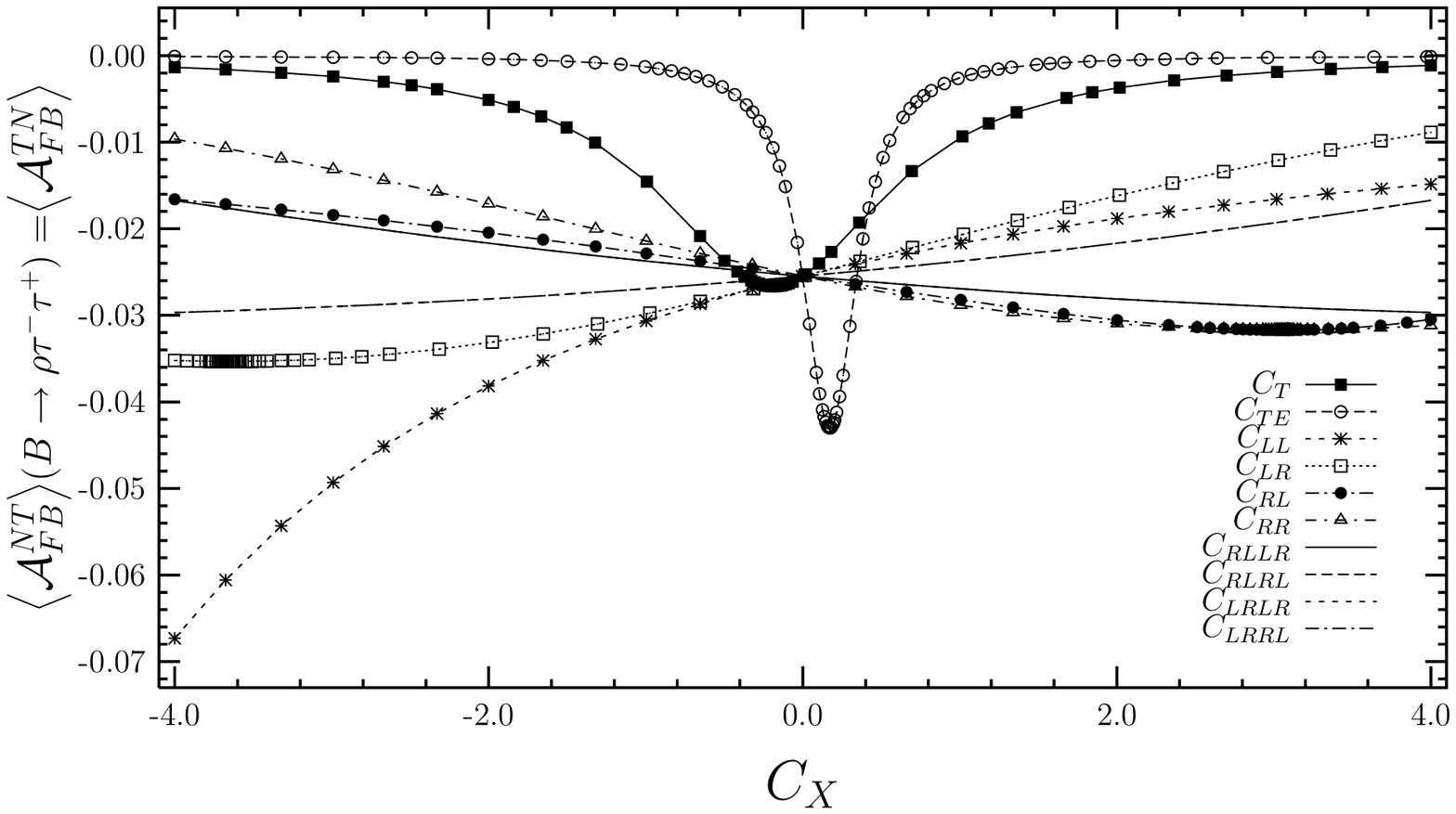}
\vskip 7.8 cm
\caption{}
\end{figure}

\begin{figure}
\vskip 1.5 cm
    \includegraphics{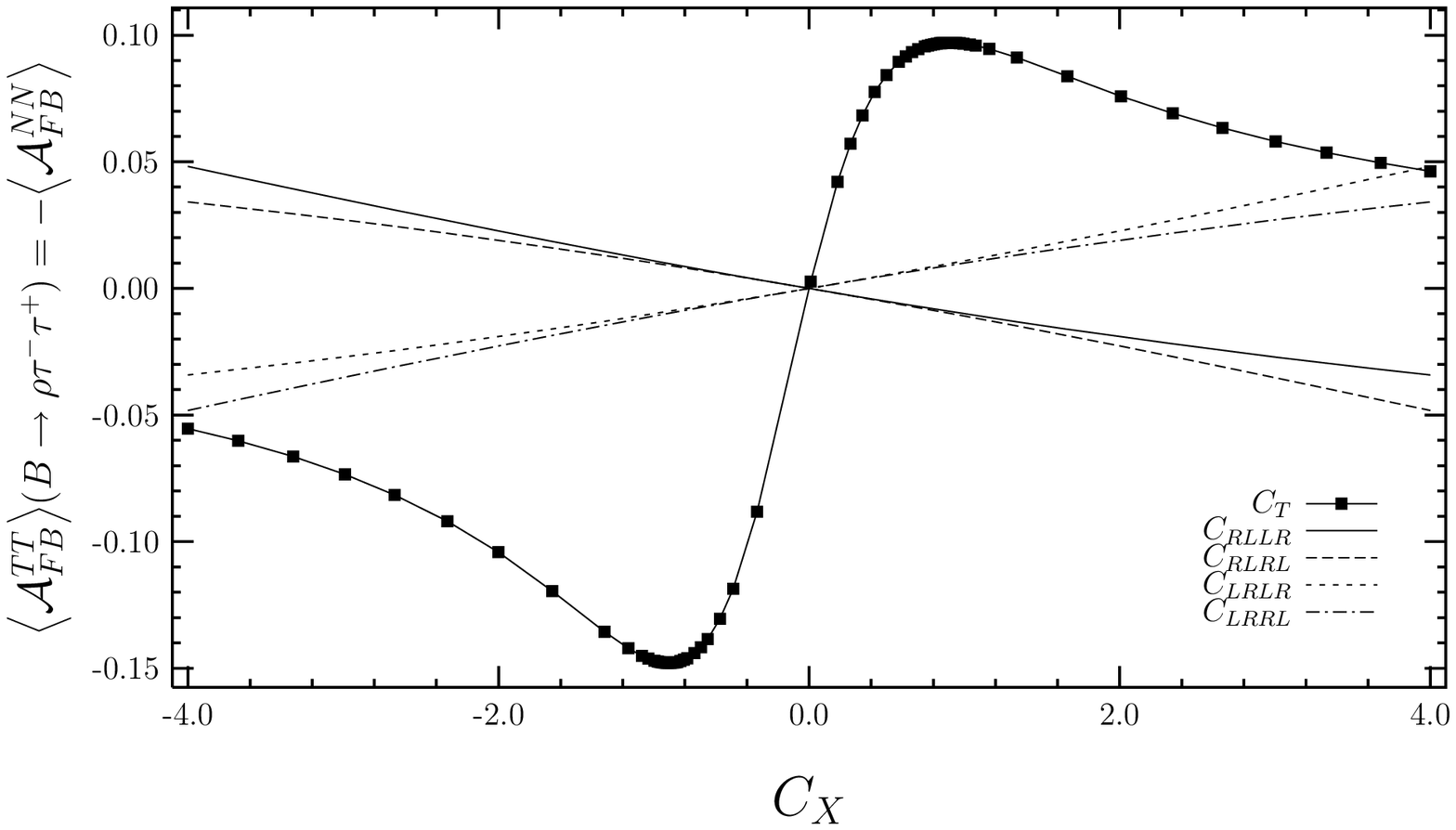}
\vskip 7.8cm
\caption{}
\end{figure}

\begin{figure}
\vskip 2.5 cm
    \includegraphics{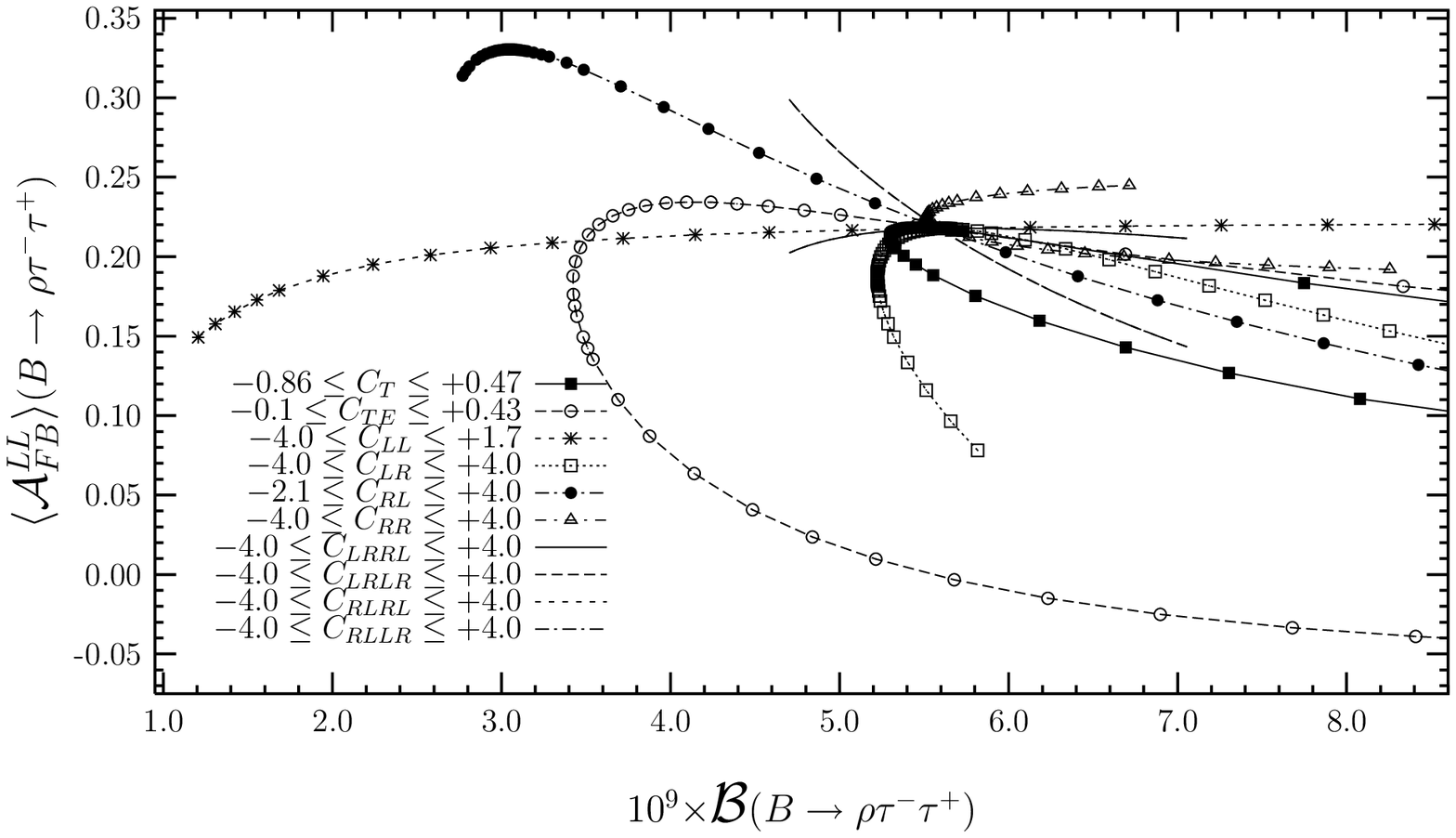}
\vskip 7.8 cm
\caption{}
\end{figure}

\begin{figure}
\vskip 1.5 cm
    \includegraphics{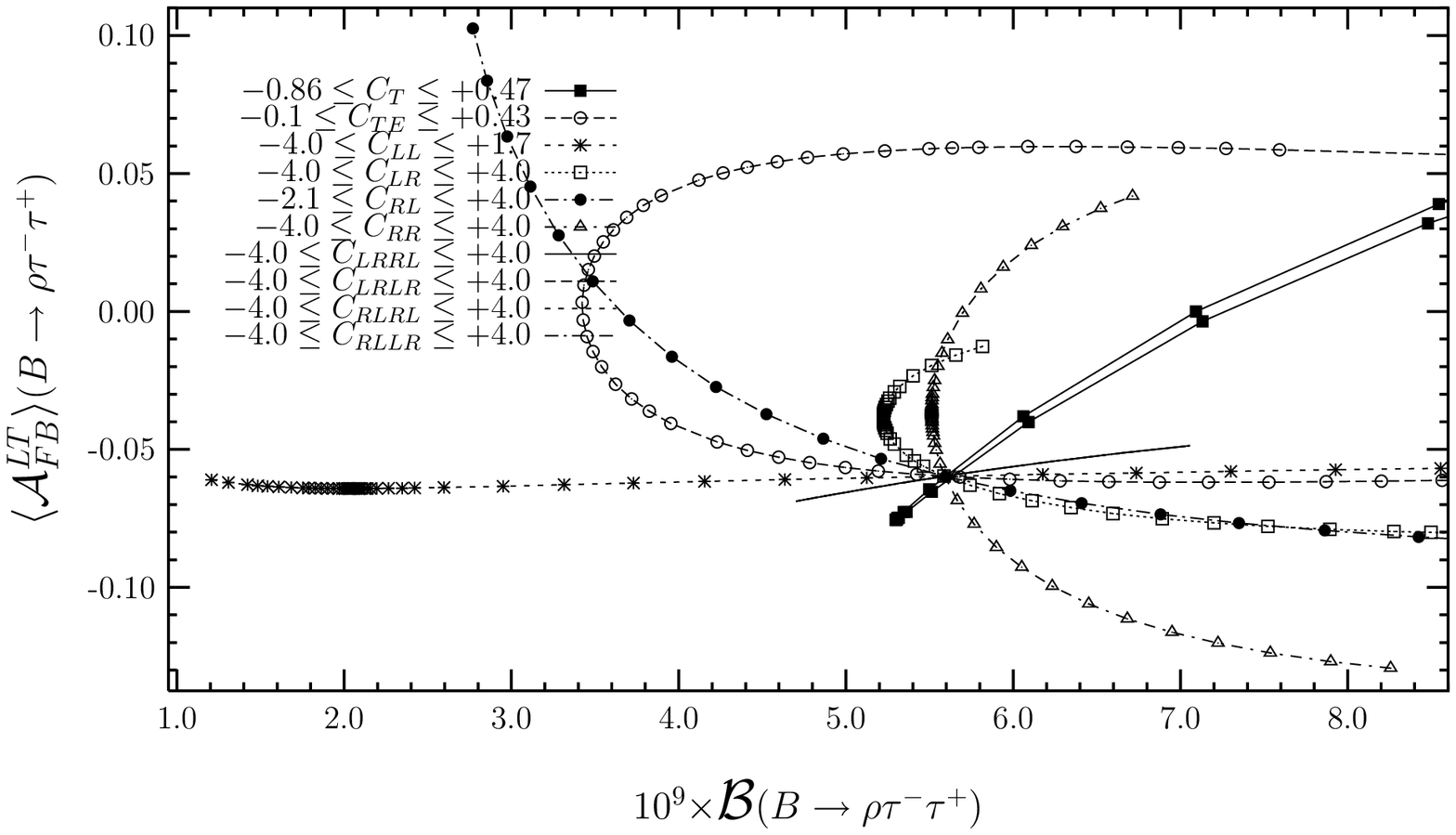}
\vskip 7.8cm
\caption{}
\end{figure}

\end{document}